\documentclass[twocolumn,preprintnumbers]{revtex4}

\usepackage{graphicx}
\usepackage{epsf}
\usepackage{amsmath,amssymb}
\usepackage{color}
\usepackage{ulem}
\newcommand{\bvec}{\boldsymbol}

\newcommand{\cent}{c}
\newcommand{\so}{so}
\newcommand{\tens}{t}

\newcommand{\ForestGreen}[1]{\textcolor[cmyk]{0.91,0,0.88,0.12}{#1}}

\begin{document}

\title{Tensor correlations in $^4$He and $^8$Be with antisymmetrized quasi cluster model}

\author{H. Matsuno, Y. Kanada-En'yo}

\affiliation{
Department of Physics, Kyoto University, Kitashirakawa Oiwake-Cho, Kyoto 606-8502, Japan
}

\author{N. Itagaki}

\affiliation{
Yukawa Institute for Theoretical Physics, Kyoto University,
Kitashirakawa Oiwake-Cho, Kyoto 606-8502, Japan
}

\begin{abstract}
In this paper,  we extend the framework of improved version of simplified method to take into account
the tensor contribution ($i$SMT) and propose AQCM-T,
tensor version of antisymmetrized quasi cluster model (AQCM). 
Although AQCM-T is phenomenological,
we can treat the $^3S$-$^3D$ coupling in the deuteron-like $T=0$ $NN$-pair induced 
by the tensor interaction
in a very simplified way,
which allows us to proceed to heavier nuclei.  
Also we propose a new effective 
interaction, V2m, 
where the triplet-even channel of the Volkov No.2 
interaction is weakened to 60\% so as to reproduce the 
binding energy of $^4\textrm{He}$ after including the tensor term of a realistic interaction. 
Using AQCM-T and the new interaction,  
the significant tensor contribution in $^4\textrm{He}$ is shown, which is
almost comparable the central interaction, where
$D$-state mixes by 8\% to the major $S$-state.
The AQCM-T model with the new interaction is also applied to $^8\textrm{Be}$.
It is found that the tensor suppression gives significant contribution 
to the short-range repulsion between two $\alpha$ clusters. 

\end{abstract}
\maketitle

\section{Introduction}
The nucleus $^4$He is the strongly bound four-nucleon system with large binding energy per nucleon 
in light mass region, and $\alpha$ particles called $\alpha$ clusters
can be basic building blocks of the nuclear structure.
Based on the assumption that nuclear systems are composed of $\alpha$ clusters,  
 $\alpha$ cluster models ~\cite{Brink,Fujiwara} have been developed and applied 
in numerous works
for the description of 
nuclear structures  including the so-called Hoyle state of $^{12}$C~\cite{Hoyle,Uegaki,THSR}.
Describing cluster states
is a challenge for the shell models 
including modern {\it ab initio} ones~\cite{Maris,Dreyfuss,Yoshida},
since quite large model space is required.
Our goal is to pave the way to generally describe the nuclear structure, both cluster and shell 
structures.
In this study, we start with the cluster side 
and construct a model that can deal with higher correlations, in particular 
the tensor correlation, in an economic way with less computational efforts.

There have been fundamental
discussions for the appearance of cluster structure in the
1960s; ``why clustering is favored?".
The appearance could be related to the nature of the meson exchange potential;
one-pion exchange potential (OPEP),
which is the exchange of isovector meson, vanishes 
when each $\alpha$ cluster has isospin $T = 0$~\cite{Shimodaya}.
As a result, intercluster interaction is weak, 
and two-pion exchange potential gives almost satisfactory 
phase shift of $\alpha$-$\alpha$ scattering.
Therefore, the appearance of cluster structure is
natural consequence of the meson theory.
In OPEP, the tensor term plays a dominant role,
thus the clustering can be considered as the embodiment of 
suppression or screening of the tensor interaction.

The tensor interaction also plays a crucial role inside $^4$He.
It has been already pointed out in {\it ab initio} calculations in 1970s 
that contribution of two-particle-two-hole (2p2h)
states is very important in $^4$He because of the strong tensor 
effect~\cite{ATMS}.
According to the modern {\it ab initio} calculations, the contribution of the tensor interaction
to the $^4$He binding energy is quite large. For instance, in the case of the AV8' potential, it is  
more than 68 MeV and even more important than the central interaction~\cite{Kamada}.
Therefore, the tensor interaction plays key roles in both  
mechanisms for the appearance of the clustering;
strong binding of each $\alpha$ cluster and weak interaction between the clusters.

It has been pointed out that this strong tensor contribution
in $^4\textrm{He}$ can be suppressed when 
another $^4\textrm{He}$ approaches, due to the Pauli blocking effect~\cite{Bando}.
The appearance of $\alpha$-$\alpha$ cluster structure in $^8\textrm{Be}$,
which is confirmed by {\it ab initio} quantum Monte Carlo calculation~\cite{QMC}, 
is also attributed to the tensor suppression effect.
In Ref.~\cite{Bando}, Brueckner theory has been introduced to estimate this 
suppression effect, while keeping each $\alpha$ cluster to a simple $(0s)^4$ configuration.
The improvement of this model space has been performed in Ref.~\cite{Yamamoto};
nevertheless it is quite important to discuss this suppression effect
by treating the tensor contribution in more direct way. 

In most of the conventional
cluster models, each $\alpha$ cluster is often assumed as 
a simple $(0s)^4$ configuration placed at some spatial point.
In such simple models, since
$\alpha$ cluster is a spin singlet object,
contributions of non-central interactions such as the tensor interaction, as well as the spin-orbit interaction, 
completely vanish, 
even though they play crucial roles in the nuclear structure.
One needs to take into account cluster breaking components to explicitly
deal with the non-central interactions.
Recently, many microscopic attempts of directly taking
into account the non-central interactions  
for the studies of cluster structure have begun.
For instance, 
the methods of antisymmetrized molecular dynamics (AMD)~\cite{KanadaEnyo:1995tb,KanadaEnyo:1995ir,AMDsupp,KanadaEn'yo:2012bj}
and Fermionic molecular dynamics (FMD)~\cite{Neff,Roth,Chernykh} combined with the 
unitary correlation method (UCOM) 
have been developed and extensively applied.
In AMD and FMD, each nucleon is independently treated as a Gaussian wave packet localized 
in the phase space,
which enables us to
describe various cluster structures and also 
the shell-model structure, where clusters are broken.
Also, the complex Gaussian centroid of the single-nucleon wave function
is suitable for taking into account the non-central interactions.
The tensor effect in $^4\textrm{He}$ has been studied by extended 
AMD~\cite{Dote:2005un}.
In UCOM, 
by  unitary transforming the Hamiltonian, 
the tensor effect is included, which 
in principle induces many-body operators up to $A$ (mass number) body,
thus the truncation of the model space is required.
Our strategy is slightly different; although it is phenomenological,
we introduce an effective model
to directly take into account the non-central interactions in a simplified manner. 

Concerning the inclusion of the rank one non-central interaction,
the spin-orbit interaction, in the cluster model, 
we proposed the antisymmetrized quasi cluster model 
(AQCM)~\cite{Simple,Masui,Yoshida2,Ne-Mg,Suhara,Suhara2015,Itagaki,Itagaki-CO,
Matsuno,Matsuno2,O24}.
By introducing a parameter for the imaginary part of the Gaussian centroids of 
  $\alpha$ clusters,
we can smoothly transform $\alpha$ clusters to
$jj$-coupling shell model wave functions, and
the transformed $\alpha$ clusters are called quasi clusters.
As it is well known,
the conventional $\alpha$ cluster models cover the model space of closure of major shells
($N=2$, $N=8$, $N=20$, {\it etc.}), but not subclosure configurations, where
the spin-orbit interaction contributes. 
Our AQCM can be regarded as an extended cluster model
that
covers also the  $jj$-coupling subclosure configurations.

However the rank two non-central
interaction, $i. e.$ the tensor interaction, is more complicated to be treated in the cluster model.
The tensor interaction has two features, the first order type and the second order type.
The first order one is rather weak and 
characterized by the attractive effect for a proton (neutron) with the 
$j$-upper orbit of the $jj$-coupling shell model and a neutron (proton) 
with $j$-lower orbit~\cite{Otsuka}, 
which can be included just by switching on the tensor interaction
using AQCM.

For the second order type (2p2h type), which is more difficult to be treated
in the cluster model,
we have proposed 
a simplified model to directly take into account the tensor contribution
(SMT)~\cite{Itagaki-SMT}. 
We started with the $(0s)^4$ configuration for an $\alpha$ cluster as an unperturbed configuration
and expressed deuteron-like excitation of a proton and a neutron to higher shells
by shifting the positions of Gaussian centroids 
of these two particles.
However, the resultant tensor contribution was not enough large as much as expected. 
Shifting the positions of Gaussian centroids
could not be sufficient in mixing higher momentum components of the 2p2h configurations.

According to the tensor optimized shell model 
(TOSM)~\cite{TOSM,TOSM2007,TOSM2009,TOSM2011,TOSM2012}
and tensor optimized AMD (TOAMD) \cite{Myo:2015rbv,Myo:2017gjc} calculations, 
the $p$ orbits of this 2p2h states
must have very shrunk shape compared with the normal shell model orbits, 
and this means that mixing of very high momentum components 
is quite important. 
Then, we further developed a improved version of SMT, which is $i$SMT~\cite{iSMT}.
In the method, imaginary parts of the Gaussian centroids are shifted.
The imaginary part of Gaussian centroid corresponds to the expectation value
of momentum for the nucleon.
The tensor interaction has the character which is suited to be described in the momentum space,
and this method is more efficient in directly mixing the
higher momentum components of 2p2h configurations.
The contribution  of the tensor interaction 
in $^4$He was more than $-40$ MeV, four times larger than the previous version.
The method was also applied to $^{16}$O,
where the tensor contribution is also large,
and this is coming from the finite size effect for the distances among 
$\alpha$ clusters with a tetrahedral configuration.
The model space of $i$SMT is further extended in high-momentum AMD
(HM-AMD)~\cite{Myo:2017amv,Myo:2017nmz},
and even more tensor contribution was obtained in $^4$He.

It should be commented that the shifting imaginary
parts of the Gaussian centroids has been already
achieved in the original AQCM for the spin-orbit force;
centroids were
shifted so that two neutrons (or two protons) in an $\alpha$
cluster have finite momenta in opposite directions.
What
is essential in $i$SMT is that high momentum component is taken into account by shifting imaginary part for
a proton and a neutron with the isospin $T = 0$.
In this sense, $i$SMT
can be regarded as an extended AQCM for the tensor effect.

In this paper, we further develop $i$SMT and newly propose AQCM-T, 
which is the tensor version of AQCM  introduced by the authors and their collaborators.
This is also regarded  
a specific version of the HM-AMD developed by
 Myo {\it et al.}. 
In the previous analyses based on $i$SMT and HM-AMD,
the tensor interaction was just added to the (conventional) effective Hamiltonian.
Since the tensor effect was already renormalized in the strong triplet-even ($^3E$) central part
of the effective Hamiltonian,
it was doubly counted.
Indeed, $^4\textrm{He}$ was too much overbound different from the 
realistic one.
In this study, we first construct the new framework of AQCM-T, and next propose a new effective 
interaction with the central and tensor parts, 
where the triplet-even part of the central interaction is weakened so as to reproduce the
binding energy of $^4$He within AQCM-T.
We analyze internal wave functions of the correlated $NN$ pairs and show 
the contribution of the tensor correlation
in relatively shorter ranges of the $^3D$ and $^3S$ channels
compared with uncorrelated ($(0s)^2$) $NN$ pair.
We also apply the method to $^8\textrm{Be}$ and study tensor effects
in two $\alpha$ cluster structure. It is found that
the
suppression of the tensor correlation significantly contributes to the short-range repulsion of two $\alpha$ clusters.

This paper is organized as follows:
In Sec.~{\bf II}, the framework,
especially for the model wave function, is explained.
In Sec.~{\bf III}, the Hamiltonian of the present model including the 
new effective interaction is described.
In sections {\bf IV} and {\bf V}, the numerical results for $^4\textrm{He}$
and $^4\textrm{Be}$ are presented, respectively. 
Summary is presented in Sec.~{\bf VI}. The fitting procedure of the tensor term of a realistic interaction
is explained in Appendix~\ref{app:a}.

\section{Formulations}  
\subsection{AQCM-T for a $NN$ pair}
In this article, we introduce a new framework called  AQCM-T.
Although all the nucleons can be treated as independent Gaussians,
we take notice of the correlation of two nucleons, which is taken into account
by properly setting the Gaussian centroids of the two nucleons.
We start the discussion with a single $NN$ pair. 

Each single-particle wave function is
written by Gaussian wave packet as 
\begin{eqnarray}
&&\psi_j(i)=\phi_{\bvec{S}_j}(\bvec{r}_i)\chi_j(s_i,\tau_i),\\
&&\phi_{\bvec{S}_j}(\bvec{r}_i)
=\left(\frac{2\nu}{\pi}\right)^{\frac{3}{4}}
e^{-\nu(\bvec{r}_i-\bvec{S}_{j})^2},
\label{spwf}
\end{eqnarray}
where 
$\bvec{S}_j$ ($j=1,2$)
is the Gaussian centroid and
$\chi_j$ is the spin-isospin wave function.
The width parameter $\nu$ is set to $\nu=0.25$ fm$^{-2}$ 
and fixed for all the cases in the present article.

For the two nucleons in a correlated $NN$ pair, 
we introduce their Gaussian centroids 
with complex conjugate values, 
\begin{equation}
\bvec{S}_{1,2}=\bvec{R}\pm \frac{i\bvec{K}}{\nu},
\end{equation}
with real vectors $\bvec{R}$ and $\bvec{K}$.
Here plus and minus singes are for $j=1$ and $j=2$, respectively,
and they corresponds to the two nucleons in the time reversal states to each other.

Then the
spatial part of the $NN$ pair wave function can be rewritten by relative and center-of-mass (cm) wave functions
using $\bvec{k}\equiv 2 \bvec{K}$ as 
\begin{eqnarray}
&&\phi_{\bvec{S}_1}(\bvec{r}_1)\phi_{\bvec{S}_2} (\bvec{r}_2)= \varphi_{\bvec{k}}(\bvec{r})  \phi_g(\bvec{r}_g),  \\
&&\varphi_{\bvec{k}}(\bvec{r})=\left(\frac{\nu}{\pi}\right)^{\frac{3}{4}}
e^{-\frac{\nu}{2}r^2+i\bvec{k}\cdot\bvec{r}
+\frac{k^2}{2\nu}},  \label{eq:phi} \\
&& \phi_g(\bvec{r}_g)
=\left(\frac{4\nu}{\pi}\right)^{\frac{3}{4}}
e^{-2\nu(\bvec{r}_g-\bvec{R})^2},
\end{eqnarray}
where $\bvec{r}=\bvec{r}_1-\bvec{r}_2$ and $\bvec{r}_g=(\bvec{r}_1+\bvec{r}_2)/2$ are the relative and cm coordinates of the $NN$ pair, respectively.
The expectation values of the coordinates and momenta are given as  
\begin{eqnarray}
&&\langle \bvec{r}\rangle =0, \qquad \langle \bvec{p}\rangle =\bvec{k}\\
&&\langle \bvec{r}_{g}\rangle =\bvec{R} , \qquad \langle \bvec{p}_{g}\rangle =0.
\end{eqnarray}
It should be noted that the Fourier components of  the relative wave function $\varphi_{\bvec{k}}(\bvec{r})$
also has a Gaussian form, which is
localized at $\bvec{k}$ with the dispersion of $\sqrt{\nu}$.

As pointed out by Myo {\it et al.},
the relative wave function $\varphi_{\bvec{k}}(\bvec{r})$ has the angler dependence 
coming from the factor $e^{i\bvec{k}\cdot\bvec{r}}$
and contains not only $S$-wave but higher partial-wave components as 
\begin{eqnarray}
e^{i\bvec{k}\cdot\bvec{r}}=4\pi\sum_{lm} i^l j_l(kr) Y_{lm}(\bvec{e}_k)
 Y_{lm}(\bvec{e}_r).
\end{eqnarray}
For the inclusion of the
tensor correlation, the $^3D$ component 
in the $T=0$ $NN$-pair,
which couples to the $^3S$ component
is essential.
Therefore, we project the $NN$ pair state on the
positive-parity state.
Suppose that $\bvec{k}$ is set along the $z$ axis as $\bvec{k}=(0,0,k)$, 
the positive-parity state $\varphi^+_k$ projected from $\varphi_{\bvec{k}}$ is 
expanded with the $l$-even basis states
as
\begin{eqnarray}
&&\varphi^+_k(\bvec{r})=
\left(\frac{\nu}{\pi}\right)^{\frac{3}{4}}
e^{-\frac{\nu}{2}r^2+\frac{k^2}{2\nu}}\cos(kz)\nonumber\\
&&=\left(\frac{\nu}{\pi}\right)^{\frac{3}{4}}
 e^{-\frac{\nu}{2}r^2+\frac{k^2}{2\nu}}
4\pi\sum_{l=\textrm{even}} \sqrt{\frac{2l+1}{4\pi}} i^l j_l(kr) Y_{l0}(\bvec{e}_r),\nonumber\\
&&=\sum_{l=\textrm{even}} 
a_l \varphi^{(l)}_k(r) Y_{l0}(\bvec{e}_r),
\label{eq:partial-NN}
\end{eqnarray}
where $a_l$ is the normalization 
factor, and 
$\varphi^{(l)}_k(r)$ is the normalized radial wave function 
of the $l$-even basis state
and
proportional to $e^{-\frac{\nu}{2}r^2} j_l(kr)$.


\subsection{AQCM-T for $^4\textrm{He}$} 

Next the $NN$ pair wave function introduced in the previous subsection
is applied to the two nucleons in $^4\textrm{He}$.

\subsubsection{Model wave function of $^4\textrm{He}$}

For the $^4\textrm{He}$ system, 
in addition to the correlated $NN$ pair introduced in the previous
subsection,
we consider a $(0s)^2$ (uncorrelated) pair, and both pairs are placed at the origin. 
The AQCM-T wave function for $^4\textrm{He}$ is expressed as 
\begin{eqnarray} \label{eq:4He}
\Phi^\textrm{AQCM-T}_{^4\textrm{He},0^+}&&=\hat{P}^{0+}{\cal A}\{\phi_{\frac{i\bvec{K}}{\nu}}\chi_1,
\phi_{-\frac{i\bvec{K}}{\nu}}\chi_2,\phi_{0}\chi_3,\phi_{0}\chi_4\}, \nonumber \\
&&
=\hat{P}^{0+}{\cal A}\{\phi_{\frac{i\bvec{K}}{\nu}}
\phi_{-\frac{i\bvec{K}}{\nu}}\phi_{0}\phi_{0}\otimes \chi_1\chi_2\chi_3\chi_4\}, \nonumber\\
\end{eqnarray}
where ${\cal A}$ is the antisymmetrizer, $\hat{P}^{0+}$ is the 
projection operator to
$J^\pi=0^+$ (in practice numerically performed), 
and $\phi_{0}=\phi_{\bvec{S}=0}$ is the 
spatial wave function for a nucleon in the
$0s$ orbit.
The spatial wave function of the total system in the intrinsic frame before the projections
is rewritten as
\begin{eqnarray}
&&\phi_{\frac{i\bvec{K}}{\nu}}
\phi_{-\frac{i\bvec{K}}{\nu}}\phi_{0}\phi_{0}
=
\phi_g(\bvec{r}_{g})\phi_g(\bvec{r}'_{g})
\varphi_{\bvec{k}}(\bvec{r})\varphi_{0}(\bvec{r}'),\label{eq:spatial} \\
&&\bvec{r}_g=\frac{\bvec{r}_1+\bvec{r}_2}{2},\qquad \bvec{r}'_g=\frac{\bvec{r}_3+\bvec{r}_4}{2},\\
&&\bvec{r}=\bvec{r}_1-\bvec{r}_2,\qquad \bvec{r}'=\bvec{r}_3-\bvec{r}_4.
\end{eqnarray}
This
means that the $NN$ correlation is taken into account through $\varphi_{\bvec{k}}(\bvec{r})$
of the correlated pair.

The AQCM-T wave function for $^4\textrm{He}$ in Eq.~\eqref{eq:4He} is a general
expression, which contains basis wave functions used in the preceding works 
by Itagaki \textit{et al.}~\cite{iSMT} and Myo \textit{et al.}~\cite{Myo:2017amv}.
In Ref.~\cite{iSMT}, the orientation of the vector $\bvec{k}$ was 
introduced along the $z$-axis, which 
is the axis of the spin quantization, and in Ref.~\cite{Myo:2017amv},
basis states with $\bvec{k}$ direction perpendicular to the $z$-axis were further introduced,
while keeping the spin orientations to the original $z$ and $-z$ directions.
In principle, 
if we prepare spin configurations properly,
the orientation of the vector $\bvec{k}$ can be
arbitrary chosen, 
because the intrinsic wave function is projected to 
the physical  $^4\textrm{He}$ state with $J=0$. 
In the present model space, 
we choose the parameter $\bvec{k}$ as $\bvec{k}=(0,0,k)$ and consider 
the important
spin and isospin configurations properly.
The present choice of the $z$ direction 
is the same as that in Ref.~\cite{iSMT}, and this is convenient when extending 
the method to heavier systems such as $^8\textrm{Be}$,
because $^4\textrm{He}$ is an axial symmetric object
in the intrinsic frame.

One should care about the redundancies originating 
from the parity and angular momentum projections as well as the Fermi statistics (antisymmetrization). 
In this model, we take into 
account all the spin and isospin configurations necessary to express 
$0^+$ states and avoid the redundancy. 
As a result, for a given $k$ value,  
the model space for $0^+$ states of $^4\textrm{He}$ contains
five independent spin and isospin configurations; 
\begin{eqnarray} \label{eq:config5}
&&\chi_1 \chi_2 \chi_3\chi_4=\nonumber \\
&&\{  
{p}\uparrow {p}\downarrow n\uparrow n\downarrow, \ \
{n}\uparrow {n}\downarrow p\uparrow p\downarrow, \ \ \nonumber \\
&& 
p\uparrow n\uparrow p\downarrow n\downarrow, \ \
p\uparrow n\downarrow p\uparrow n\downarrow, \ \ \nonumber \\
&& p\uparrow n\downarrow p\downarrow n\uparrow
\}.
\end{eqnarray}
Owing to the projection to
$J^\pi=0^+$, 
$\Phi^\textrm{AQCM-T}_{^4\textrm{He},0^+}$
contains only the  $S$-wave ($\varphi^{(0)}_k$) and 
$D$-wave ($\varphi^{(2)}_k$) components,
which are coupled with the total intrinsic spin $S=0$ and $S=2$ of four nucleons, respectively.
Here $\varphi^{(l)}_{k}$ stands for
the $l$-wave relative wave function for the $NN$ pair in the partial wave expansion of Eq.~\eqref{eq:partial-NN}.
Note that $\varphi^{(0)}_{k=0}$ expresses
the uncorrelated $NN$ pair with the $(0s)^2$ configuration.

When we ignore small breaking of the isospin symmetry by the Coulomb interaction, the five configurations in Eq.~\eqref{eq:config5} can be
reduced into three channels with respect to spin and isospin symmetries of 
the $NN$ pair as 
\begin{eqnarray}
\label{eq:channel3-1}
^1S:&& \phi_g(\bvec{r}_{g})\phi_g(\bvec{r}'_{g})\otimes
\varphi^{(0)}_k(r)\varphi^{(0)}_{0}(r')   \nonumber\\
&&
\otimes Y_{00}(\bvec{e}_r) Y_{00}(\bvec{e}_{r'})  \otimes  
\chi^\sigma_{0} \chi^\sigma_{0}
\otimes [ \chi^\tau_{1} \chi^\tau_{1}]_{T=0},
\end{eqnarray}
\begin{eqnarray}
\label{eq:channel3-2}
^3S: && \phi_g(\bvec{r}_{g})\phi_g(\bvec{r}'_{g})\otimes
\varphi^{(0)}_k(r)\varphi^{(0)}_{0}(r')   \nonumber\\
&&\otimes
Y_{00}(\bvec{e}_r) Y_{00}(\bvec{e}_{r'})  \otimes [\chi^\sigma_{1}  
\chi^\sigma_{1}]_{S=0}\otimes
\chi^\tau_{0} \chi^\tau_{0},
\end{eqnarray}
\begin{eqnarray}
\label{eq:channel3-3}
^3D:&& \phi_g(\bvec{r}_{g})\phi_g(\bvec{r}'_{g})\otimes
\varphi^{(2)}_k(r)\varphi^{(0)}_{0}(r')   \nonumber\\
&&\otimes\left[ Y_{20}(\bvec{e}_r) Y_{00}(\bvec{e}_{r'})  \otimes
  [\chi^\sigma_{1} \chi^\sigma_{1}]_{S=2}\right]_{J=0}\otimes
\chi^\tau_{0} \chi^\tau_{0},\nonumber\\
\end{eqnarray}
where $\chi^\sigma_S$ ($\chi^\tau_T$) is the spin (isospin) function of the $NN$ 
pairs coupled to 
the spin $S$ (isospin $T$) state.
The first (second) configuration 
in \eqref{eq:channel3-1} (\eqref{eq:channel3-2})
takes into account the $NN$ correlation in the $^1S$ ($^3S$) channel.
In principle, the
short-range correlation caused by the repulsive hard core
contributes in these channels, 
and amplitudes of two nucleons close to each other 
wave should be suppressed;
however the central interaction adopted in the present study 
is not a realistic nuclear force but an effective interaction without a hard core.
The third configuration is the so-called $D$-state component, which is essential 
in the
tensor correlation. 
We call the first, second, and third configurations, the 
$^1S$, $^3S$, and $^3D$ channels, respectively.  

In the present framework,
$\Phi^{\textrm{AQCM-T}}_{^4\textrm{He},0^+}$ defined in Eqs.~\eqref{eq:4He} and \eqref{eq:spatial}
is 
a basis wave function
specified by the $k$ value 
in Eq.~\eqref{eq:phi} and the spin-isospin configuration.
The total wave function  for the ground state,
$\Psi_{^4\textrm{He},\textrm{gs}}$, 
is expressed by linear combination of 
various $k$ values and the spin and isospin configurations as 
\begin{eqnarray}
\label{eq:gcm-4He}
\Psi_{^4\textrm{He},\textrm{gs}}=c_0
\Phi^{0s}_{^4\textrm{He}}+
 \sum_{k} \sum_\beta c(k,\beta) \Phi^{\textrm{AQCM-T}}_{^4\textrm{He},0^+}(k,\beta),
\end{eqnarray} 
where $\beta$ is the label for the spin-isospin configurations 
in Eq.~\eqref{eq:config5}
(or channels in (\ref{eq:channel3-1})-(\ref{eq:channel3-3})).
Here, the first term of
$\Phi^{0s}_{^4\textrm{He}}$ is the $(0s)^4$ wave function equivalent to  
$\Phi^{\textrm{AQCM-T}}_{^4\textrm{He},0^+}(k,\beta)$ with $k=0$ and 
$\beta=p\uparrow p\downarrow n\uparrow n\downarrow$. 
The coefficients $c_0$ and $c(k,\beta)$ are determined by diagonalizing the 
norm and Hamiltonian matrices comprised of the basis wave functions.
The superposition 
with respect to $k$ 
in Eq.~\eqref{eq:gcm-4He}
is nothing but the
expansion of the correlated $NN$ pair wave function 
in terms of Gaussians with mean momentum $\bvec{k}$
in the momentum space, and the sum of $\beta$ corresponds to the 
coupled-channel calculation of $\beta=\{^1S$, $^3S$, and $^3D\}$.

In the present framework, we 
take notice on a single $NN$ pair  among the four nucleons
and explicitly treat
the two-body correlations,
but we omit higher order correlations, where more than two
nucleons are involved. 
This ansatz is supported by the four-body calculations by Horii 
\textit{et al.} in Ref.~\cite{Horii:2011nf}, which demonstrates that the 
$D$-state coupling with the $S$-state, which is dominant,
in a single $NN$ pair with $T=0$ is essential in describing the $^4\textrm{He}$ ground state. 
This is a natural consequence of the bosonic feature of two  $NN$ pairs 
with $T=0$
in $^4\textrm{He}$.

{In this article, we present a new framework and call ``AQCM-T'',
because this is a tensor version of the AQCM, in which 
clusters are changed into quasi clusters characterized 
by the complex Gaussian centroids. 
The AQCM has been originally proposed to describe the breaking of
$nn$ and $\alpha$
clusters by the spin-orbit interaction at the nuclear surface,
and this
can be regarded as an extended version of the Brink cluster model 
or a specific version of the AMD model.
The AQCM treatment of introducing the imaginary part
for the Guassian centroids
has been applied to a $pn$ pair to 
describe the tensor correlation in $^4\textrm{He}$ by Itagaki and Tohsaki in Ref.~\cite{iSMT}, 
in which the method was called  ``$i$SMT''. 
The model space of $i$SMT was extended in ``HM-AMD''  by Myo \textit{et al.} for the study of
the tensor correlations of $^4\textrm{He}$
 in  Ref.~\cite{Myo:2017amv}. 
In order to treat short-range correlations as well as the tensor correlations,
they have achieved further extension of the  HM-AMD model 
by taking into account higher-order correlations beyond two-body \cite{Myo:2017nmz}.
Our
parameter $\bvec{k}$ 
in Eq.~\eqref{eq:phi}
for the imaginary centroids of the Gaussian wave packets 
is related to
the notations of the parameters $\bvec{d}$ in $i$SMT and 
$\bvec{D}$ in HM-AMD  as
$\bvec{d}=\bvec{D}=\bvec{K}/{\nu}=\bvec{k}/(2\nu) $.
It should be commented that the model spaces of Refs.~\cite{iSMT,Myo:2017amv}
correspond to subsets of the present spin and isospin configurations  
defined in Eq.~\eqref{eq:config5}. 
One of the key points of the present model is that 
we explicitly represent not only  
the isospin symmetry of the correlated pair but also that of the $(0s)^2$ pair, which is essential 
in describing the isoscalar property of the $^4\textrm{He}$ ground state. 

\subsubsection{Parameter settings for $^4\textrm{He}$}
For the ground state of $^4\textrm{He}$
($^4\textrm{He}_\textrm{gs}$), 
we perform calculations with the three channels 
defined in 
\eqref{eq:channel3-1}-\eqref{eq:channel3-3}
($\beta=\{^1S,^3S, ^3D\}$).
This three-channel calculation can be practically done using 
five configurations defined in Eq.~\eqref{eq:config5}.
If we can omit the effect of the charge symmetry breaking by the Coulomb interaction, 
these two sets of configurations are 
equivalent. 
Indeed, the three-channel 
calculation gives almost the same result as that of full five configurations, indicating that the
symmetry breaking in the isospin space is negligibly small. 
For each channel,  
the basis states with
$k=0.5,1.0,\ldots, 5.5$ fm$^{-1}$ (11 points) are 
adopted
in addition to the $(0s)^4$ configuration. As a result,  
the total 
number of the basis states
in Eq.~\eqref{eq:gcm-4He} 
is $11\times 3+1=34$
corresponding to the dimension of the Hamiltonian to be diagonalized.
We also perform calculations with truncated model space 
and compare with the full result
to 
clarify the roles of the $^1S$, $^3S$, and $^3D$ components. 

\subsection{AQCM-T for $^{8}\textrm{Be}$}
\subsubsection{AQCM-T wave function of $2\alpha$}
Our aim is to investigate the tensor effect in heavier nuclei.
Here we extend the AQCM-T framework to $^8\textrm{Be}$ 
with two $\alpha$ cluster structure, 
in which 
one of $\alpha$ clusters is changed from the $(0s)^4$ configuration to
the correlated $^4\textrm{He}$ wave function
previously explained.
We  label the correlated $\alpha$ cluster as $\alpha_k$, and another $\alpha$ cluster 
with
the $(0s)^4$ configuration is labeled as $\alpha_0$.
We place $\alpha_k$ at $\bvec{R}=\frac{\bvec{d}_\alpha}{2}$ and 
$\alpha_0$ at $\bvec{R}'=-\frac{\bvec{d}_\alpha}{2}$ 
with the relative distance of $\bvec{d}_\alpha$.
After the antisymmetrization, 
the $0^+$ projected $2\alpha$  wave function is
\begin{equation}
\Phi^{\textrm{AQCM-T}}_{2\alpha,0^+}(k,\beta, \bvec{d}_\alpha)
=\hat{P}^{0+}{\cal A}\left\{ 
\Phi_{\alpha_k}(k,\beta,\bvec{R}) 
\Phi_{\alpha_0}(\bvec{R}') 
\right\},
\label{eq:8Be}
\end{equation}
where $\beta$ is the label for the spin and isospin configurations of the 
$\alpha_k$ cluster.  
The two $\alpha$ clusters are expressed using 
the AQCM-T wave function for $^4$He as
\begin{eqnarray}
&&\Phi_{\alpha_k}(k,\beta,\bvec{R})=
\Phi^\textrm{AQCM-T}_{^4\textrm{He},+}(k,\beta,\bvec{R})\nonumber \\
&&=\frac{1+\hat{P}_k}{2}{\cal A}\{\phi_{\bvec{R}+\frac{i\bvec{K}}{\nu}}
\phi_{\bvec{R}-\frac{i\bvec{K}}{\nu}}
\phi_{\bvec{R}}\phi_{\bvec{R}}
\otimes \chi_1\chi_2\chi_3\chi_4 \}, \nonumber \\
\ \\
&& \Phi_{\alpha_0}(\bvec{R}') = \Phi^{0s}_{^4\textrm{He}} (\bvec{R}') \nonumber \\
&&={\cal A}\{\phi_{\bvec{R}'}\phi_{\bvec{R}'}\phi_{\bvec{R}'}\phi_{\bvec{R}'}
\otimes {p\uparrow}{p\downarrow} {n\uparrow} {n\downarrow} \}.
\end{eqnarray}
Here $\bvec{K}=(0,0, k/2)$,
and the operator $\hat{P}_k$ transforms the imaginary part of
the correlated $NN$ pair as $k\to -k$. Thus, the
intrinsic wave function of the correlated $NN$ pair is 
projected onto the positive-parity state by the operator $(1+\hat{P}_k)/2$.
The parameter $\bvec{d}_\alpha$ for the relative distance is chosen as 
$\bvec{d}_\alpha=(d_\alpha \sin\theta_\alpha, 0, d_\alpha \cos\theta_\alpha)$.
For fixed $d_\alpha$, states with
various $k$, $\theta_\alpha$, and $\beta$ values are superposed
as
\begin{eqnarray}\label{eq:gcm-8Be}
&&\Psi_{2\alpha,0^+}(d_\alpha)=c_0
\Phi^{\textrm{BB}}_{2\alpha,0^+}(d_\alpha)\nonumber\\
&&+
\sum_{k,\beta,\theta_\alpha} c(k,\beta,\theta_\alpha) 
\Phi^{\textrm{AQCM-T}}_{2\alpha,0^+}(k,\beta,\bvec{d}_\alpha),
\end{eqnarray}
where $\Phi^{\textrm{BB}}_{2\alpha,0^+}$ is Brink-Bloch (BB)
$2\alpha$ cluster wave function projected to $0^+$, 
\begin{eqnarray}
&&\Phi^{\textrm{BB}}_{2\alpha,0^+}(d_\alpha)=\hat{P}^{0+}{\cal A}\left\{ 
\Phi_{\alpha_0}(\bvec{R}) 
\Phi_{\alpha_0}(\bvec{R}') 
\right\}.
\end{eqnarray}
For each $d_\alpha$ value, 
the coefficients $c_0$ and $c(k,\beta,\theta_\alpha)$ are determined 
by diagonalizing the 
norm and Hamiltonian matrices.
We investigate 
the tensor correlations of the two $\alpha$ system as a function of 
$d_\alpha$.

In the present framework, not only for the
relative motion between clusters, the angular momentum of 
the subsystem ${\alpha_k}$ is practically projected;
although the projection in $\Psi_{2\alpha,0^+}(d_\alpha)$ is only
for the total angular momentum,
the double projection is achieved. 
This is
owing to the rotational symmetry of the 
$\alpha_0$ cluster, the axial symmetry of the 
$\alpha_k$, and superposition effect of states with various 
$\theta_\alpha$ values ($0\le \theta_\alpha \le \pi/2$).
The range of $\pi/2 \le \theta_\alpha \le \pi$ is redundant in the present
case, since the intrinsic parity of the 
$\alpha_k$ cluster is already projected. 

Here the angle $\theta_\alpha$ is treated as a generator coordinate, 
while
the parameter $d_\alpha$ is fixed, and the inter-cluster wave function 
is localized around $d_\alpha$.
In principle, $d_\alpha$ can be also treated as 
a generator coordinate; superposing $\Psi_{2\alpha,0^+}(d_\alpha)$
with different $d_\alpha$ values gives better solution for the 
inter-cluster motion.  However, such calculation
requires huge computational costs, and we perform our calculation for each 
fixed $d_\alpha$ value. 

In the present AQCM-T framework for the two $\alpha$ system, 
we explicitly treat the $NN$ correlation in   
one of the two $\alpha$ clusters, but we omit configurations that 
$NN$ pairs in both $\alpha$ clusters are simultaneously excited from $(0s)^2$, 
which could significantly contribute
in the asymptotic region ($d_\alpha \to \infty$). 
If each $^4\textrm{He}_\textrm{gs}$ cluster  
contains the
$(0s)^4$ component ($|0s\rangle$) 
still dominantly
and the mixing of the correlated component 
($|\textrm{corr}\rangle$) 
is minor in amplitude 
as 
$|^4\textrm{He}_\textrm{gs}\rangle =(1-|\varepsilon|^2) |0s\rangle + \varepsilon |\textrm{corr}\rangle$
with small enough $\varepsilon$, 
the present ansatz is a good approximation within the order 
of ${\cal O}(\varepsilon)$, owing to the bosonic symmetry for the exchange
of two $\alpha$'s.

\subsubsection{Parameter setting for $^8\textrm{Be}$}
For the generator coordinate $\theta_\alpha$ for the angle, we adopt
five mesh points of $\theta_\alpha=0, \pi/8, \ldots, 
\pi/2$, which gives almost converged results.
Regarding the correlated $\alpha$ cluster ($\alpha_k$) wave function, 
we truncate the configurations introduced for $^4$He
in order to save the computational costs; 
here we employ only two channels of $\beta=\{^3S$, $^3D$\}, 
because the $^1S$ channel is found to be not 
essential for the tensor correlations in $^4\textrm{He}$ as we discuss later.
The calculation with these two channels is practically performed by
employing the following 
three configurations, 
\begin{eqnarray} \label{eq:config3}
&&\chi_1 \chi_2 \chi_3\chi_4\nonumber \\
&&=
\{  
p\uparrow n\uparrow p\downarrow n\downarrow, \ \ 
p\uparrow n\downarrow p\uparrow n\downarrow, \ \ 
p\uparrow n\downarrow p\downarrow n\uparrow
\}.\nonumber
\end{eqnarray}
For the parameter $k$ in Eq.~\eqref{eq:8Be}, we use three points $k=\{1,2,3\}$ fm$^{-1}$, 
which efficiently describes the properties of $^4\textrm{He}_\textrm{gs}$.
Therefore, the number of the basis states in Eq.~\eqref{eq:gcm-8Be} 
corresponding to the dimension of the diagonalization
is $3\times 2 \times 5+1=31$ for a given distance of $d_\alpha$.

\subsection{$0s$, ${}^1S$, ${}^3S$, and ${}^3D$ probabilities}
In this study, we analyze the probabilities of
the ${}^1S,{}^3S,{}^3D$ components in the 
obtained $^4$He and two $\alpha$ states ($|\Psi\rangle$),
\begin{eqnarray}
{\cal P}_{{}^1S,{}^3S,{}^3D}=|\langle \Psi|
\hat{P}_{{}^1S,{}^3S,{}^3D} |\Psi \rangle|,
\end{eqnarray}
and the $0s$ probability is given as 
\begin{eqnarray}
{\cal P}_{0s}=|\langle 0s|\Psi\rangle |^2.
\end{eqnarray}
Here 
$|\Psi \rangle=|\Psi_{^4\textrm{He},\textrm{gs}} \rangle$ and 
$|0s\rangle=|\Phi^{0s}_{^4\textrm{He}}\rangle$ for the $^4$He system,
and 
$|\Psi \rangle=|\Psi_{2\alpha,0^+}\rangle$
and 
$|0s\rangle=|\Phi^{\textrm{BB}}_{2\alpha,0^+}\rangle$ 
for the two $\alpha$ system.
$\hat{P}_{{}^1S,{}^3S,{}^3D}$ are the projection operators onto the 
${}^1S,{}^3S,{}^3D$ components.
We also 
calculate the probabilities 
\begin{eqnarray}
{\cal P}^\perp_{{}^1S,{}^3S}&=&|\langle \Psi|\Lambda^\perp_{0s}
\hat{P}_{{}^1S,{}^3S}\Lambda^\perp_{0s} |\Psi \rangle|,\\
\Lambda^\perp_{0s}&=&1-|0s\rangle \langle 0s|, 
\end{eqnarray}
of the correlated ${}^1S,{}^3S$ components, which are defined in the 
$\Lambda^\perp_{0s}$-projected space orthogonal to the $0s$ state.
Note that ${\cal P}^\perp_{{}^1S,{}^3S}$ somewhat depends on the adopted
width parameter $\nu$ of the Gaussian wave packet defined in Eq.~\eqref{spwf}, and therefore, 
one should be careful in quantitative 
discussions on the absolute values of ${\cal P}^\perp_{{}^1S,{}^3S}$.

\section{Hamiltonian}  
The Hamiltonian used in the present calculation is
\begin{align}
\hat{H}&=\sum_{i}^A
\hat{T}_i-\hat{T}_\mathrm{G} \nonumber \\
&
+\sum_{i<j}^A
\left[
\hat{V}_\mathrm{\cent}(i,j)+\hat{V}_\mathrm{\so}(i,j)+\hat{V}_\mathrm{\tens}(i,j)+\hat{V}_\mathrm{Coulomb}(i,j)\right],
\end{align}
where $\hat{T}_i$ is the kinetic energy operator of $i$th nucleon,
and the total kinetic energy operator for the cm motion ($\hat{T}_\mathrm{G}$)
is subtracted.
The two-body interaction consists of central interaction
($\hat{V}_\mathrm{\cent}$), spin-orbit interaction ($\hat{V}_\mathrm{\so}$),
tensor interaction ($\hat{V}_\mathrm{\tens}$), and 
Coulomb interaction ($\hat{V}_\mathrm{Coulomb}$) terms.
The Coulomb interaction for the protons is approximated by a seven-range Gaussian form.

\subsection{Central interaction}

For the central interaction $\hat{V}_\mathrm{\cent}$, 
we use an effective nucleon-nucleon interaction. Our central interaction is based on the
Volkov No.2~\cite{Volkov}, which is a phenomenological one and reproduces
the $\alpha$-$\alpha$ scattering phase shift when the Majorana exchange parameter is properly chosen. The original Volkov interaction has only the Wigner and Majorana exchange terms,
but here we add
Bartlett and Heisenberg terms as
\begin{align}
\hat{V}_\mathrm{\cent}=&\left[V_\alpha\exp\left(-\frac{r_{ij}^2}{\alpha^2}\right)+V_\rho\exp\left(-\frac{r_{ij}^2}{\rho^2}\right)\right] \nonumber \\
&\times\left[w+b\hat{P}^\sigma_{ij}-h\hat{P}^\tau_{ij}-m\hat{P}^\sigma_{ij}\hat{P}^\tau_{ij}\right],
\end{align}
where $V_\alpha=-60.65\,\mathrm{MeV}$, $V_\rho=61.14\,\mathrm{MeV}$, $\alpha=1.80\,\mathrm{fm}$, and $\rho=1.01\,\mathrm{fm}$, which are the original values.

This is a phenomenological interaction, and tensor effect as well as the hard-core contribution is effectively renormalized 
in the central interaction, and if we just add the tensor interaction to the Volkov interaction, the tensor effect 
is doubly counted.
As explained in subsection \ref{sec:newint},
we introduce 
a new effective interaction containing the central and tensor interaction
terms by modifying the original Volkov No.2 interaction.

\subsection{Spin-orbit interaction}
For the spin-orbit interaction $\hat{V}_\mathrm{\so}$, we use the spin-orbit part of 
the G3RS interaction~\cite{G3RS}, which is a realistic nucleon-nucleon interaction, given by
\begin{align}
\hat{V}_\mathrm{\so}=&\left[u_1\exp\left(-\frac{r_{ij}^2}{\eta_1^2}\right)+u_2\exp\left(-\frac{r_{ij}^2}{\eta_2^2}\right)\right] \nonumber \\
&\times\hat{P}_{ij}(^3O)\hat{\bvec{L}}_{ij}\cdot\hat{\bvec{S}}_{ij},
\end{align}
where $u_1=600\,\mathrm{MeV}$, $u_2=-1050\,\mathrm{MeV}$, $\eta_1=0.447\,\mathrm{fm}$ and $\eta_2=0.6\,\mathrm{fm}$\, which are the values of ``case 1" of G3RS.
Here $\hat{P}_{ij}(^3O)$ is the projection operator to the
triplet odd ($^3O$) state.

\subsection{Tensor interaction} 
\label{subsec:tensor}
For the tensor interaction $\hat{V}_\mathrm{\tens}$, we introduce 
three difference ones and compare the results. The first one is
the tensor part of G3RS~\cite{G3RS}, which is a realistic interaction
and its spin-orbit part
was explained in the previous subsection, given by

\begin{align}
\hat{V}_\mathrm{\tens}^{\mathrm{(G3RS)}}=&\hat{S}_{ij}\times
\nonumber\\
&\left[  \sum_{n=1}^3V_{\mathrm{\tens},n}^{\mathrm{(G3RS)}^3E}\hat{P}_{ij}(^3E)\exp\left(-\frac{r_{ij}^2}{\eta_{\mathrm{\tens},n}^2}\right)\right. \nonumber\\
&\left.+\sum_{n=1}^3V_{\mathrm{\tens},n}^{\mathrm{(G3RS)}^3O}\hat{P}_{ij}(^3O)\exp\left(-\frac{r_{ij}^2}{\eta_{\mathrm{\tens},n}^2}\right)\right], \\
\hat{S}_{ij}=&3(\hat{\bvec{\sigma}}_i\cdot\hat{\bvec{r}}_{ij})(\hat{\bvec{\sigma}}_j\cdot\hat{\bvec{r}}_{ij})/r_{ij}^2-(\hat{\bvec{\sigma}}_i\cdot\hat{\bvec{\sigma}}_j),
\label{G3RS-T}
\end{align}
where $\hat{P}_{ij}(^3E)$ is the projection operator to the
triplet even ($^3E$) state.
We use the parameter set of ``case 1" of G3RS.

The second one is the
Furutani tensor interaction~\cite{Furutani}, which is constructed based on the G3RS tensor part
but gives stronger tensor contribution than the G3RS,
given by
\begin{align}
\hat{V}_\mathrm{\tens}^\mathrm{(Furutani)}=&\hat{S}_{ij}\times \nonumber\\
&
\sum_{n=1}^3V_{\mathrm{\tens},n}^{\mathrm{(Furutani)}}(W_n-H_n\hat{P}^\tau_{ij})r_{ij}^2\exp\left(-\beta_nr_{ij}^2\right).
\label{Furutani-T}
\end{align}
This interaction was used in our previous SMT and $i$SMT works. 
Compared with the Gaussian form of the G3RS tensor part,
the Furutani tensor has the $r^2$-weighted Gaussian form,
which allows us to calculate the matrix element easily, when 
local Gaussian type of the wave function is introduced as in the present case.

The third one is again the G3RS tensor part, but 
newly constructed by fitting the G3RS tensor part 
with the $r^2$-weighted Gaussian form,

\begin{align}
\hat{V}_\mathrm{\tens}^\textrm{(3R-fit)}=&\hat{S}_{ij}\times
\nonumber\\
&\left[ 
\sum_{n=1}^{n_\textrm{max}=3}
V_{\mathrm{\tens},n}^{\textrm{(3R-fit)}^3E}\hat{P}_{ij}(^3E)
r_{ij}^2\exp\left(-\beta_nr_{ij}^2\right)\right. \nonumber\\
&+\left. 
\sum_{n=1}^{n_\textrm{max}=3}
V_{\mathrm{\tens},n}^{\textrm{(3R-fit)}^3O}\hat{P}_{ij}(^3O)
r_{ij}^2\exp\left(-\beta_nr_{ij}^2\right)\right]. 
\label{New-T}
\end{align}

It has the $r^2$-weighted Gaussian form with 3 ranges the same as
the Furutani tensor interaction. This
form can be easily adopted in the present framework.
In the present work,  we fit the G3RS tensor part using this functional form
and propose a new G3RS-like tensor interaction in a convenient form. The
details of the fitting are explained in Appendix  \ref{app:a}.
The parameter sets of all three tensor interactions are summarized 
in Table~\ref{table-tensor}.
The radial part of the $^3E$ and $^3O$ components of the
G3RS tensor part, Furutani tensor, and 
the new 3-range fit (3R-fit) of G3RS tensor part
are compared in Fig.~\ref{app.fig.TensorCompare}.  \par

\begin{table}[ht]
\caption{The parameter sets of G3RS tensor part (case 1), Furutani tensor, and new 3-range fit (3R-fit)
tensor, which is G3RS tensor part fitted by using the functional form of the Furutani tensor
defined in Eqs.~\eqref{G3RS-T}-\eqref{New-T}.
}
\label{table-tensor}
\centering
\begin{tabular}{rrrr}
\hline \hline
\multicolumn{4}{c}{G3RS tensor part (case 1)} \\ \hline
$n$ & $1$ & $2$ & $3$ \\ \hline
$\eta_{\mathrm{\tens},n}\,\mathrm{(fm})$ & $2.5$ & $1.2$ & $0.447$ \\
$V_{\mathrm{\tens},n}^{\mathrm{(G3RS)}^3E}\,\mathrm{(MeV})$ & $-7.5$ & $-67.5$ & $67.5$ \\
$V_{\mathrm{\tens},n}^{\mathrm{(G3RS)}^3O}\,\mathrm{(MeV})$ & $2.5$ & $20$ & $-20$ \\ \hline \hline
\multicolumn{4}{c}{Furutani tensor} \\ \hline
$n$ & $1$ & $2$ & $3$ \\ \hline
$\beta_n\,\mathrm{(fm^{-2}})$ & $0.53$ & $1.92$ & $8.95$ \\
$V_{\mathrm{\tens},n}^{\mathrm{(Furutani)}}\,\mathrm{(MeV\cdot{}fm^{-2}})$ & $-16.96$ & $-369.5$ & $1688.0$ \\
$W_n$ & $0.3277$ & $0.4102$ & $0.5$ \\
$H_n$ & $0.6723$ & $0.5898$ & $0.5$ \\ \hline \hline
\multicolumn{4}{c}{New 3-range fit tensor} \\ \hline
$n$ & $1$ & $2$ & $3$ \\ \hline
$\beta_n\,\mathrm{(fm^{-2}})$ & $0.53$ & $1.92$ & $8.95$ \\
$V_{\mathrm{\tens},n}^{\textrm{(3R-fit)}^3E}\,\mathrm{(MeV\cdot{}fm^{-2})}$ & $-17.02$ & $-209.89$ & $-289.59$ \\
$V_{\mathrm{\tens},n}^{\textrm{(3R-fit)}^3O}\,\mathrm{(MeV\cdot{}fm^{-2})}$ & $5.27$ & $62.91$ & $89.87$ \\ \hline \hline
\end{tabular}
\end{table}

\begin{figure}[htbp]
\centering
\includegraphics[width=.5\textwidth, trim= 0pt 0pt 0pt 0pt, clip]{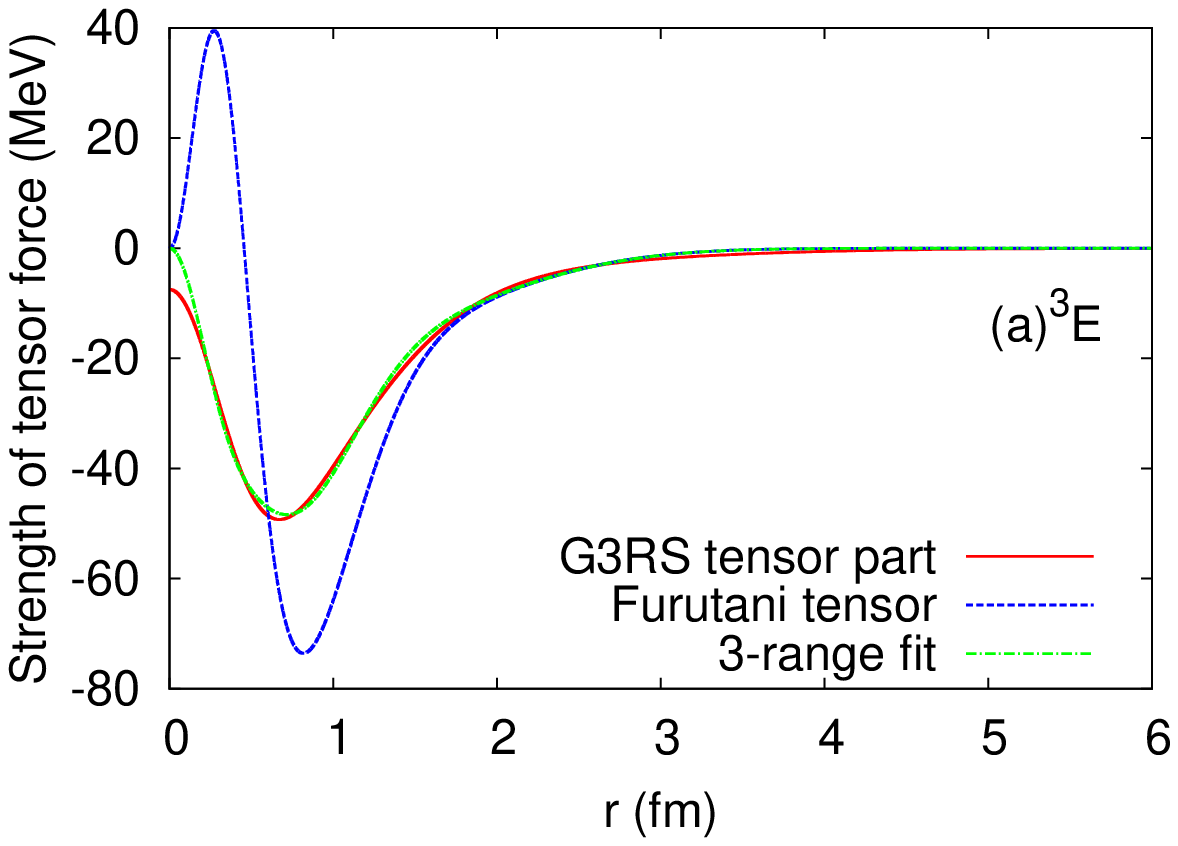}
\includegraphics[width=.5\textwidth, trim= 0pt 0pt 0pt 0pt, clip]{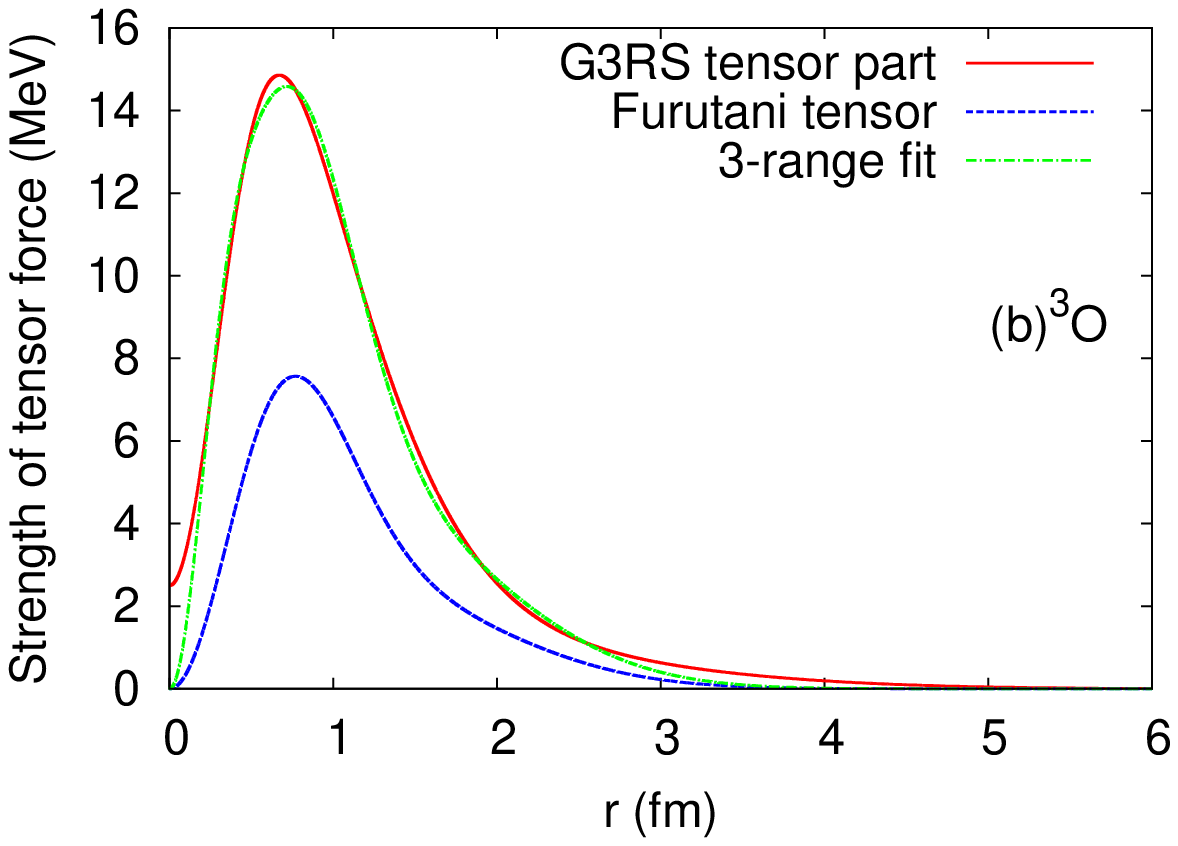}
\caption{The comparison of the radial part of the G3RS tensor term (solid line), 
Furutani tensor (dotted line), and 3-range fit (dash-dotted line).
((a): triplet even part, (b): triplet odd part).}
\label{app.fig.TensorCompare}
\end{figure}

\subsection{New interactions and parametrization}
\label{sec:newint}


In this paper,
we compare the results of different parameter sets
for the central and tensor interactions, 
which give major contributions to the binding energy, 
while the spin-orbit and Coulomb parts are fixed. 

As mentioned previously, 
the Volkov interaction is a phenomenological  central interaction, and
the tensor contribution is effectively renormalized. 
Therefore,
if we just add the tensor interaction to the Volkov interaction, the tensor effect is doubly counted.
In this study, we introduce a modified version of the Volkov interaction and 
use it  for the central part of the  new effective interaction containing the central and tensor 
terms.
We design the new interaction so as to reasonably reproduce energies of $^4$He and two-nucleon systems.

We start with the Volkov No.2 interaction with $m=0.6$, $w=1-m$, $b=h=0.15$.
This parameter set is called ``V2''
often used in the conventional cluster models.
This set has been known to 
reproduces  the
energy and radius of $^4$He$_\mathrm{gs}$, and also the $\alpha$-$\alpha$ scattering 
phase shift within the $(0s)^4$ configuration for the $\alpha$ cluster(s).
The Bartlett and Heisenberg parameters of  $b=h=0.15$ are chosen so as to 
reproduce the $NN$ scattering lengths of $^1S$ and $^3S$ without the tensor interaction as
$a_s=-24$ and $a_t=5.4$ fm, respectively (the experimental values are 
$a_s=-18.5\pm  0.4$ fm~\cite{SCnn} and $a_s=-23.749\pm  0.008$ fm~\cite{SCpn} for the $nn$ and $pn$ channels, respectively,
and $a_t=5.423\pm 0.005$ fm~\cite{SCnn}). 

\begin{table}[ht]
\caption{The parameter sets of the different combinations of the tensor and central 
interactions, and properties of two nucleon systems.
The column $V_\mathrm{\cent}$ is for the central interactions, 
and the column $V_\mathrm{\tens}$ is for the tensor interactions. 
The V2m interaction 
is the modified version of Volkov No.2 newly introduced in the present study.
The experimental data of the $^1S$ scattering length ($a_s$) is
$a_s=-23.749\pm  0.008$ fm for the $pn$ channel~\cite{SCpn} 
($a_s=-18.5\pm  0.4$ fm for the $nn$ channel~\cite{SCnn}), and that of the
deuteron binding energy ($-\epsilon_d$) is $-\epsilon_d=2.22$ MeV.
}
\label{tab.ParameterSetsOfTheInteraction}
\centering
\begin{tabular}{ccccc}
\hline \hline
& V2m-3R & V2-3R & V2-F &V2   \\ \hline
$V_\mathrm{\cent}$ & V2m & V2 & V2 & V2 \\
$V_\mathrm{\tens}$ & 3R-fit & 3R-fit & Furutani & $-$ \\ \hline
$a_s\,(\mathrm{fm})$ & $-24$ & $-24$ & $-24$ & $-24$ \\
$\epsilon_d\,(\mathrm{MeV})$ & $-4.38$&$-11.02$ & $-18.46$ & $-2.65$ \\ \hline\hline
\end{tabular}
\end{table}

Now we  construct a new effective interaction 
by combining the newly modified V2 interaction for the
central part and the 3-range fitted G3RS interaction for the tensor part.
In the original V2 interaction, 
the large tensor contribution 
is effectively renormalized in the $^3E$ central term. As a result, 
the $^3E$ part of the V2 interaction is much stronger than 
the $^1E$ part, as the ratio of $^3E/^1E=1.3/0.7$, 
inconsistent to the realistic interactions. 
To avoid double counting of the tensor contribution, we
reduce the $^3E$ part by introducing a factor $\delta_{^3E}$ as 
\begin{align}
\hat{V}^\mathrm{(V2m)}_\mathrm{\cent}=\left(1-(1-\delta_{^3E})\hat{P}_{ij}(^3E)\right)\hat{V}^\mathrm{(V2)}_\mathrm{\cent},
\end{align}
 where $\hat{V}^\mathrm{(V2)}_\mathrm{\cent}$ is the original V2. 
We adopt $\delta_{^3E}=0.6$ (reduction  of the $^3E$ part to 60\% of the original strength) 
which gives reasonable binding energy of $^4$He$_\mathrm{gs}$
within AQCM-T after including the tensor interaction.
This modified central interaction is labeled as ``V2m''. 
After the reduction, the  $^3E$  strength  
becomes almost the same as $^1E$ one with the ratio of $^3E/^1E=0.78/0.7$.
Then we add the 3-range fitted G3RS tensor interaction. 
We label the newly constructed interaction containing the central 
and tensor interactions as ``V2m-3R''.

We also introduce other two interactions 
by just adding tensor interactions to the 
V2 interaction without any reduction
 and compare the results with that obtained by the V2m-3R interaction.
One is the ``V2-3R'' interaction,  in which the 3-range fitted G3RS tensor interaction
is added to the V2 interaction. 
The other is ``V2-F'', where the Furutani tensor 
interaction is added to V2.

The parameter sets for these four interactions (V2m-3R, V2-3R,V2-F, and V2)
are summarized in Table~\ref{tab.ParameterSetsOfTheInteraction}.
The $^1S$ scattering length
 and the deuteron binding energies 
obtained with these interactions are also shown.
It should be commented that our newly constructed interaction, V2m-3R,
gives reasonable results for the low-energy properties 
of both $^1E$ and $^3E$ channels, 
whereas the V2 interaction combined with the tensor interactions has the overbinding problem of the deuteron,
because of the double counting of the tensor effect in the $^3E$ channel.

\section{Results of $^4\textrm{He}$}

\subsection{Properties of $^4\textrm{He}$} 
Properties of 
$^4\textrm{He}_\textrm{gs}$ obtained with  
AQCM-T and V2m-3R, V2-3R, and V2-F interactions
are shown in Table~\ref{tab:energy}. The total energy ($E$), contributions of the
kinetic term ($T$), central ($V_\mathrm{\cent}$) and tensor ($V_\mathrm{\tens}$) interactions,  
root-mean-square (rms) matter radii ($R_m$), 
and $0s$, ${}^1S$, ${}^3S$, and ${}^3D$ probabilities are listed.
The result calculated with the 
single $(0s)^4$ configuration ($\Phi^{0s}_{^4\textrm{He}}$)  using the 
V2 interaction is also shown for comparison.

\begin{table}[t]
\caption{Energies, radii, and probabilities of $^4\textrm{He}$
obtained with AQCM-T full configurations and 
the V2m-3R, V2-3R, and V2-F interactions together with the 
experimental energy and radius \cite{angeli13}. 
The result for the $(0s)^4$ state
with the V2 interaction is also shown (V2:$(0s)^4$).
In V2m-G3RS, V2m is used for the central part,
and the precise (20-range) fit of G3RS tensor part is used. See Appendix A
for the precision of the 20-range fit.
}
\label{tab:energy}
\begin{center}
\begin{tabular}{crrrrrr}
\hline
\hline
	&	V2m-3R	& V2-3R	& 	V2-F	 & V2:$(0s)^4$ & V2m-G3RS  & exp.\\ \hline
$E$ (MeV)	&$	-30.3 	$&$	-52.6 	$&$	-69.2 	$&$	-27.9 	$&$	-30.7 	$&$	-28.296	$\\
$T$ (MeV)	&$	64.6 	$&$	72.3 	$&$	86.1 	$&$	46.7 	$&$	64.9 	$&$		$\\
$V_\mathrm{\cent}$ (MeV)	&$	-56.7 	$&$	-83.3 	$&$	-85.1 	$&$	-75.3 	$&$	-56.7 	$&$		$\\
$V_\mathrm{\tens}$ (MeV)	&$	-39.9 	$&$	-43.2 	$&$	-72.2 	$&$	0.0 	$&$	-40.6 	$&$		$\\
$R_m$ (fm) 	&$	1.46 	$&$	1.38 	$&$	1.33 	$&$	1.50 	$&$	1.46 	$&$	1.455	$\\
${\cal P}_{0s}$	&$	0.901 	$&$	0.867 	$&$	0.801 	$&$	1.00 	$&$	0.899 	$&$		$\\
${\cal P}_{{}^3D}$	&$	0.077 	$&$	0.082 	$&$	0.112 	$&$- 	$&$	0.079 	$&$		$\\
${\cal P}^\perp_{{}^3S}$	&$	0.018 	$&$	0.050 	$&$	0.086 	$&$- 	$&$	0.019 	$&$		$\\
${\cal P}^\perp_{{}^1S}$	&$	0.004 	$&$	0.016 	$&$	0.027 	$&$	- 	$&$	0.004 	$&$		$\\
\hline
\hline	
\end{tabular}
\end{center}
\end{table}

For V2-3R (V2-F), 
$^4\textrm{He}$ is unrealistically overbound as seen in 
much larger binding energy of $-E=52.6$ MeV ($-E=69.2$ MeV)  and the 
smaller radius of $R_m=1.38$ fm ($R_m=1.33$ fm) 
compared with the experimental values of 
$-E=28.296$ MeV and $R_m=1.455$ fm,
because of extra attraction by the strong tensor interaction
(tensor effect is already renormalized in the V2 interaction).

On the contrary, the V2m-3R interaction gives reasonable binding energy 
of $-E=30.3$ MeV,
because the $^3E$ central term is reduced to 60\% of the V2 interaction. 
In the present paper, we use this
V2m-3R as the default parameter set of the interaction, though
it is possible to fine tune the reduction factor to exactly reproduce
the experimental binding energy.
In practical calculations of  heavier systems, possible truncations of the model space may be 
required
to save computational costs. 
Therefore, this reduction factor for the  $^3E$ central term 
can be regarded as an adjustable parameter, which may depend on the  
model space adopted. 

It is quite instructive to
compare the contribution of each term of the Hamiltonian obtained 
in two different cases;
AQCM-T with V2m-3R and $(0s)^4$ configuration with 
V2; the latter is the nuclear interaction containing only the central part.
The total energy is almost the same; however, 
the contributions of $T$, $V_\mathrm{\cent}$, and $V_\mathrm{\tens}$  are much different. 
The contribution of the central interaction is reduced by $\sim$20 MeV
in V2m-3R,
because of the weaker $^3E$ central interaction
compared with that in V2. 
The remarkable feature of V2m-3R is that
large gain of the tensor energy compensates this reduction
and even overcomes the increase of
the kinetic energy.
It should be stressed that
this effect is
attributed to the $D$-state mixing with the dominant $S$-state component.
Although the $D$-state mixing is only 8\%,
the second order perturbation causes
significant gain of the tensor energy through  the 
$^3S$-$^3D$ coupling.

The AQCM-T calculation with V2m-3R gives the radius of 
$R_m=1.46$ fm, which well agrees with the experimental rms point-proton radius, 
$1.455$ fm, reduced from the observed charge radius. 
The matter density distribution is shown in 
Fig.~\ref{fig:rho} together with the single Gaussian shape 
(the 
$(0s)^4$ configuration with $\nu=0.264$ fm$^{-2}$ that gives the equivalent radius of $R_m=1.46$ fm).
About 10\% enhancement of the
 central density is obtained in the AQCM-T calculation, 
because of the $NN$ correlations beyond the simple $(0s)^4$ 
configuration. 

In Table~\ref{tab:energy}, we also show
the result of the G3RS tensor interaction combined with the 
V2m interaction (labeled as ``V2m-G3RS''), which are practically calculated 
by using the precise (20-range) fit of the G3RS tensor part. One can see 
that the 3-range fit used in V2m-3R gives almost equivalent 
contribution of each term of the Hamiltonian compared with 20-range fit.

\begin{figure}[t]
\begin{center}
\includegraphics[width=6cm]{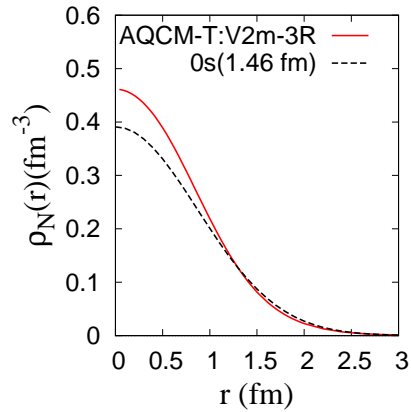} 
\end{center}
  \caption{\label{fig:rho}
Matter density distribution of $^4\textrm{He}_\textrm{gs}$ obtained with AQCM-T and V2m-3R.
The single Gaussian shape for the 
$(0s)^4$ configuration with $\nu=0.264$ fm$^{-2}$
that gives the equivalent radius 1.46 fm is also drawn.
}
\end{figure}

\subsection{$NN$ correlations in $^4\textrm{He}$}

\subsubsection{Contributions of correlated $NN$ pairs}
Next we discuss the $NN$ correlations in $^4\textrm{He}$, which are incorporated 
in the present AQCM-T calculation
by introducing the 
correlated $NN$ pairs.
Figure \ref{fig:phi-k} (a)
shows the squared overlap (${\cal O}_\beta$) 
of $\Psi_{^4\textrm{He},\textrm{gs}}$ with each basis state of AQCM-T  
specified by $k$ and $\beta$ shown in Eq.~\eqref{eq:gcm-4He}. The overlap with the ${}^3D$ channel
 is calculated as 
\begin{eqnarray}
{\cal O}_{{}^3D}(k)=
|\langle \Phi^{\textrm{AQCM-T}}_{^4\textrm{He},0^+}(k,{}^3D)|
\Psi_{^4\textrm{He},\textrm{gs}}
\rangle|^2, 
\end{eqnarray}
The overlaps with the ${}^{1,3}S$ channels are 
defined for the space orthogonal to $|0s\rangle$ as
\begin{eqnarray}
{\cal O}_{{}^{1,3}S}(k)=
|\langle \Phi^{\textrm{AQCM-T}}_{^4\textrm{He},0^+}(k,{}^{1,3}S)\Lambda^\perp_{0s}|
\Psi_{^4\textrm{He},\textrm{gs}}
\rangle|^2.
\end{eqnarray}
This is to measure the correlated $^{1,3}S$ components beyond 
the simple $(0s)^2$ pair.
The ground state 
($\Psi_{^4\textrm{He},\textrm{gs}}$)
has the largest overlap with the 
correlated $^3D$ pair at $k\sim 1.5-2.0$ fm$^{-1}$,
indicating that the intermediate momentum dominantly contributes to the tensor
correlation.
The present result is qualitatively consistent with
the result of  Ref.~\cite{Myo:2017amv}, in which the interactions are almost equivalent to  
V2-F of the present paper.
But quantitatively speaking, 
the result with V2m-3R is more or less different from that with  
V2-F; in the latter case the dominant contribution shifts to
slightly higher region of $k$, around 
$k\sim 2$ fm$^{-1}$. This interaction gives unrealistically overbound $^4\textrm{He}$,
because the tensor interaction is already renormalized in the central part of V2-F.
It is worth mentioning that $\Psi_{^4\textrm{He},\textrm{gs}}$ has finite overlap with the correlated $^3S$ pair, which 
gives non-negligible contribution to the tensor correlation, as discussed later. 

In order to clarify the contribution and role of 
each channel and basis state, we perform the AQCM-T calculations 
within truncated model spaces.
At first, we truncate the channels, $\beta$=$\{^1S, ^3S,^3D\}$, which is 
the truncation of the spin-isospin space;
we omit the $\{^1S\}$ and $\{^1S, ^3S\}$ channel(s) and 
perform two- and single-channel calculations 
using only $\beta=\{^3S,^3D\}$ and $\{^3D\}$ channel(s), respectively,
whereas we employ all the basis states for the $k$ values in Eq.~\eqref{eq:gcm-4He}.
In Table \ref{tab:truncation}, the results of two- and single-channel calculations with the V2m-3R interaction are listed and
compared with the those of the three-channel ($\beta=\{^1S, ^3S,^3D\}$) calculation.
The two-channel calculation gives quite similar result to the full (three-channel)
one, 
indicating there is almost no effect of the $^1S$ correlation. 
However, if we compare
the two-channel and single-channel calculations, it can be seen that 
the $^3S$ truncation gives significant effects on the $T$, $V_\mathrm{\cent}$, and $V_\mathrm{\tens}$ energies,
even if it gives minor effect on the total energy $E$.
For instance, 
the $V_\mathrm{\tens}$ contribution is suppressed by about 5 MeV when the 
correlated ${}^3S$ component is missing, because it  
directly couples with the  $NN$ pair in the $^3D$ state with $T=0$.
Although the ${}^3D$ component plays a primary role 
in the tensor correlation,
the coupling of the
two channels, ${}^3S$ and ${}^3D$, is necessary to 
quantitatively describe the features of the tensor correlation.
The single-channel calculation only describes basic features  of 
$^4\textrm{He}$, such as
tensor contribution in energy or $D$-state probability, qualitatively.

For comparison, we also show the 
results of three-, two-, and single-channel calculations obtained 
with the V2-F interaction in Table \ref{tab:truncation-vf}.
Unlike the V2m-3R case, the inclusion of the correlated $^3S$ component 
significantly contributes to all energy terms ($E$, $T$, $V_\mathrm{\cent}$, and $V_\mathrm{\tens}$) as well as the $D$-state probability.
However, 
it may be an artifact because of the unrealistic overbinding of $^4\textrm{He}$ 
due to the double counting of the tensor contribution in the central and tensor terms.

\begin{figure}[t]
\begin{center}
\includegraphics[width=8.5cm]{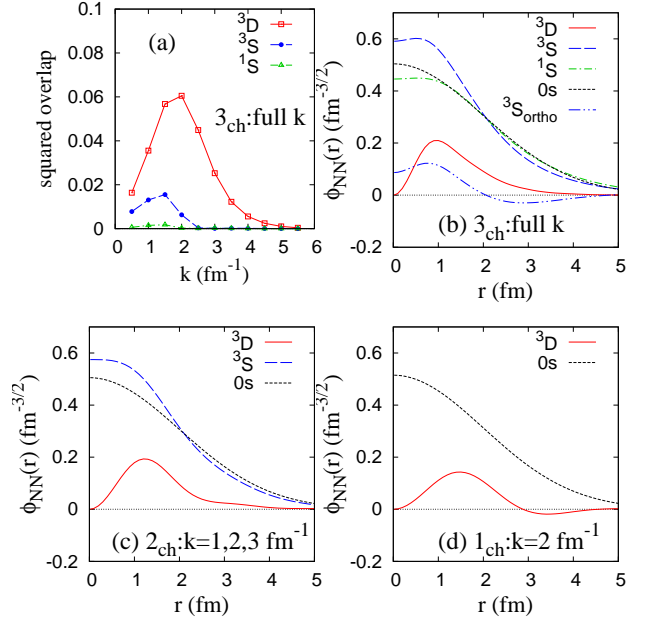} 
\end{center}
  \caption{	\label{fig:phi-k}
The squared overlaps  ${\cal O}_\beta(k)$ and pair wave functions 
$\phi_{NN}(r)$ in $^4\textrm{He}_\textrm{gs}$  calculated with AQCM-T and the V2m-3R interaction. 
(a): Squared overlaps obtained by the full calculation (three-channel calculation with full $k$ configurations,
 $k=\{0.5, 1.0, \ldots, 5.5\}$ fm$^{-1}$),
(b): pair wave functions in the $^3D$, $^3S$, and $^1S$, and $0s$ components 
obtained by the full calculation, (c): those obtained by the two-channel calculation with $k=\{1,2,3\}$ fm$^{-1}$, 
(d): those obtained by the 
single-channel calculation with $k=2$ fm$^{-1}$.
In the panel (b), the $^3S$ pair wave functions in the $(0s)^4$ configuration 
and that ($^3S_\textrm{ortho}$)  in the orthogonal configuration 
$(1-|0s\rangle \langle 0s|)|\Psi_{^4\textrm{He},\textrm{gs}}\rangle$ are also shown.
}
\end{figure}

\begin{table*}[t]
\caption{Energies, radii, and probabilities of $^4\textrm{He}$
obtained with the truncated configurations using 
the V2m-3R interaction. The results obtained by 
three-, two-, single-channel calculations  
with full $k$ configurations ($k=\{0.5, 1.0, \ldots, 5.5\}$ fm$^{-1}$), the two-channel calculation with 
$k=\{1,2,3\}$ fm$^{-1}$, and single-channel calculation with 
$k=2\ \textrm{fm}^{-1}$ are shown together with 
the result for the $(0s)^4$ state.
\label{tab:truncation}
}
\begin{center}
\begin{tabular}{crrrrrr}
\hline
\hline	
　　& 3$_\textrm{ch}$	&	2$_\textrm{ch}$ &	1$_\textrm{ch} $	&	$2_\textrm{ch}$ & 1$_\textrm{ch}$ & $(0s)^4$	\\
 $\beta$ &  $\{^1S,^3S,^3D\}$ & $\{^3S,^3D\}$ & $\{^3D\}$ &  $\{^3S,^3D\}$ & $\{^3D\}$ &  \\
$k$ &	full	&	full	 &	full		&	$\{1,2,3\}$ & $\{2\}$ & \\ \hline
$E$ (MeV)	&$	-30.3 	$&$	-30.0 	$&$	-28.2 	$&$	-28.2 	$&$	-23.1 	$&$	-8.3 	$\\	
$T$ (MeV)	&$	64.6 	$&$	65.7 	$&$	58.9 	$&$	64.8 	$&$	56.4 	$&$	46.7 	$\\	
$V_\mathrm{\cent}$ (MeV)	&$	-56.7 	$&$	-57.1 	$&$	-53.7 	$&$	-57.4 	$&$	-54.0 	$&$	-55.8 	$\\	
$V_\mathrm{\tens}$ (MeV)	&$	-39.9 	$&$	-40.2 	$&$	-35.0 	$&$	-37.3 	$&$	-26.9 	$&$	0.0 	$\\	
$R_m$ (fm) &$	1.46 	$&$	1.44 	$&$	1.49 	$&$	1.43 	$&$	1.50 $& 1.50 	\\	
${\cal P}_{0s}$&$	0.901 	$&$	0.903 	$&$	0.927 	$&$	0.905 	$&$	0.940 	$&$	1.00	$\\	
${\cal P}_{{}^3D}$	&$	0.077 	$&$	0.078 	$&$	0.073 	$&$	0.080 	$&$	0.060 	$&$	-	$\\	
${\cal P}^\perp_{{}^3S}$&$	0.018 	$&$	0.019 	$&$	- 	$&$	0.016 	$&$	- 	$&$	-	$\\	
${\cal P}^\perp_{{}^1S}$	&$	0.004 	$&$	-	$&$	- 	$&$	- 	$&$	- 	$&$	-	$\\	
\hline
\hline
\end{tabular}
\end{center}
\end{table*}

\begin{table}[ht]
\caption{Energies, radii, and probabilities of $^4\textrm{He}$
obtained with truncated configurations using 
the V2-F interaction. The results of 
three-, two-, single-channel calculations with 
full $k$ configurations are listed.
\label{tab:truncation-vf}
}
\begin{center}
\begin{tabular}{crrrr}
\hline
\hline	
	&	3$_\textrm{ch}$	&	2$_\textrm{ch}$ &	1$_\textrm{ch} $	\\
 $\beta$ &  $\{^1S,^3S,^3D\}$ & $\{^3S,^3D\}$ & $\{^3D\}$ \\
$k$	&	full	&	full	 &	full	\\ \hline
$E$ (MeV)	&$	-69.2 	$&$	-68.9 	$&$	-60.3 	$\\	
$T$ (MeV)	&$	86.1 	$&$	85.5 	$&$	65.1 	$\\	
$V_\mathrm{\cent}$ (MeV)	&$	-85.1 	$&$	-84.4 	$&$	-73.1 	$\\	
$V_\mathrm{\tens}$ (MeV)	&$	-72.2 	$&$	-72.1 	$&$	-54.2 	$\\	
$R_m$ (fm) 	&$	1.33 	$&$	1.33 	$&$	1.49 	$\\	
${\cal P}_{0s}$	&$	0.801 	$&$	0.803 	$&$	0.902 	$\\	
${\cal P}_{{}^3D}$ &$	0.112 	$&$	0.112 	$&$	0.098 	$\\	
${\cal P}^\perp_{{}^3S}$&$	0.086 	$&$	0.085 	$&$	- 	$\\	
${\cal P}^\perp_{{}^1S}$ &$	0.027 	$&$	- 	$&$	- 	$\\	
\hline
\hline
\end{tabular}
\end{center}
\end{table}

Next, we truncate the $k$ values in Eq.~\eqref{eq:gcm-4He}; we perform the two-channel 
($\beta=\{^3S,^3D\}$)
calculations with reduced 
number of the basis states with different $k$ values.
Here we employ only three values of  $k=\{1,2,3\}$ fm$^{-1}$ for the ${}^3S$ and ${}^3D$ channels, which represent the important features 
of the ground state of $^4\textrm{He}$ and cover most of the functional space,
as one can see in Fig.~\ref{fig:phi-k} (a) the overlap with the full calculation. 
As expected, the two-channel calculation only with 
$k=\{1,2,3\}$ fm$^{-1}$ efficiently describes the properties of  $^4\textrm{He}$ 
in the level almost
comparable to the full calculation. 
On the other hand, when we further reduce the model space and 
perform the single-channel calculation with a single $k=2$ fm$^{-1}$ configuration, 
we obtain the binding energy of $-E=23.1$ MeV. 
This energy is much lower
compared with 
$-E=8.3$ MeV of the pure $(0s)^4$ case owing to the
mixing of the single correlated configuration;
however, compared with the full calculation, 
the $V_\mathrm{\tens}$ contribution is significantly reduced, 
indicating that superposition of different $k$ configurations in the 
$^3S$ and $^3D$ channels is important to quantitatively describe the tensor correlation.

\subsubsection{Pair wave functions}

Using the partial wave expansion  of 
$\varphi^+_k(\bvec{r})$ shown in Eq.~\eqref{eq:partial-NN},
we reconstruct the intrinsic wave function of the correlated $NN$ pair,
which we call the pair wave function $\phi_{NN}(r)$.
The pair wave functions 
$\phi_{NN}(r)$ defined here are those for the $NN$ pair with correlations in the 
${}^{1}S$, ${}^{3}S$, and ${}^3D$ components
of $\Psi_{^4\textrm{He},\textrm{gs}}$ as, 
\begin{equation}  
{}^{1}S:\phi_{NN}(r)\varphi^{(0)}_{0}(r')  \otimes Y_{00} Y_{00}  \otimes \chi^\sigma_{0} \chi^\sigma_{0}
\otimes [ \chi^\tau_{1} \chi^\tau_{1}]_{T=0},\nonumber
\end{equation}
\begin{equation}  
{}^{3}S:\phi_{NN}(r)\varphi^{(0)}_{0}(r') \otimes Y_{00} Y_{00} \otimes [\chi^\sigma_{1} \chi^\sigma_{1}]_{S=0}\otimes
\chi^\tau_{0} \chi^\tau_{0}, \nonumber
\end{equation}
and 
\begin{eqnarray}  
{}^{3}D:\phi_{NN}(r)\varphi^{(0)}_{0}(r') \otimes \left[ Y_{20} Y_{00}  \otimes 
 [\chi^\sigma_{1} \chi^\sigma_{1}]_{S=2}\right]_{J=0}\otimes
\chi^\tau_{0} \chi^\tau_{0},\nonumber
\end{eqnarray}
 respectively.
They are given by the linear combination of $\varphi^{(0)}_k(r)$ or $\varphi^{(2)}_k(r)$, respectively,
and their Fourier transformation is related to the overlap ${\cal O}(k)$ of the 
corresponding channel. 
Figure~\ref{fig:phi-k} (b) shows
the pair wave functions $\phi_{NN}(r)$ in
the ground state
($\Psi_{^4\textrm{He},\textrm{gs}}$)
obtained with the full AQCM-T basis states and the V2m-3R interaction.
The ${}^3S$ pair wave function 
in the  $(0s)^4$ configuration projected from the ground state  
($|0s\rangle \langle 0s|\Psi_{^4\textrm{He},\textrm{gs}}\rangle$)
and that in the orthogonal (correlated component 
($(1-|0s\rangle \langle 0s|)|\Psi_{^4\textrm{He},\textrm{gs}}\rangle$) 
are also shown.
The pair wave function in the $^3D$ component has a peak around 
$r\sim 1$ fm region and shows a tail behavior in the $2\lesssim r \lesssim 3$ fm region.
The amplitude at the peak in the short distances is represented by
high $k$ components, whereas the long-range tail 
is expressed by low $k$ components. It is also interesting to see
that the $^3S$ pair wave function 
shows a significant enhancement around $r\sim 1$ fm,  consistent with the
peak position of the $^3D$ pair wave function. The enhancement of the 
$^3S$ pair wave function in this region is caused by 
the $^3S$-$^3D$ coupling attributed to the tensor interaction, 
which effectively provides an extra attraction for the $^3S$ channel. 
This $^3S$-$^3D$ coupling gives the answer why mixing of 
the correlated $^3S$ component has significant effect on the tensor correlation
in $^4\textrm{He}$, discussed previously.

For more quantitative discussion on the spatial extent of the $^3D$ pair, 
we calculate the rms distance of the pair wave function defined as
\begin{equation}
r_\textrm{pair}\equiv \sqrt{\int dr r^4 |\phi_{NN}(r)|^2 /\int dr r^2 |\phi_{NN}(r)|^2}.
\end{equation}
We obtain $r_\textrm{pair}=1.70$ fm for the $^3D$ pair, which is smaller
than $r_\textrm{pair}=2.24$ fm for the $^3S$ pair
(for the $^3S$ pair in the pure $(0s)^4$ state, $r_\textrm{pair}=\sqrt{3/(2\nu)}=2.45$ fm).

Since the enhancement of the $^3D$ and $^3S$ pair wave functions 
are seen in the $r\lesssim 2$ region, we can say that this region 
is of special importance 
for the $T=0$ pair because of the tensor correlation. 
This region of $r\lesssim 2$ for the $NN$ pair roughly corresponds 
to the internal area of $r_i\lesssim 1$ fm for the total $^4\textrm{He}$ system.

Let us turn to the pair wave functions ($\phi_{NN}(r)$)  
obtained by using truncated model space. 
Fig.~\ref{fig:phi-k} (c) shows $\phi_{NN}(r)$ for 
the two-channel ($\beta=\{^3S,^3D\}$) calculation with $k=\{1,2,3\}$ fm$^{-1}$, 
and Fig.~\ref{fig:phi-k} (d) shows that for the single-channel ($\beta={}^3D$) calculation with $k=2$ fm$^{-1}$.
In the two-channel calculation with $k=\{1,2,3\}$ fm$^{-1}$, 
$\phi_{NN}(r)$ shows similar behaviors to that of the full calculation, that is, 
the appearance of
short-range peak and long-range tail for $^3D$ and 
short-range enhancement for $^3S$.
On the other hand, in the single-channel calculation 
with $k=2$ fm$^{-1}$, somewhat different features 
of the pair wave function are seen.
The $^3D$ pair wave function shows a short-range peak, but it is milder and slightly shifted toward the outer region, 
$r\sim 1.5$ fm, than that of the full calculation. Moreover, in the long distance region, 
the  pair wave function has a negative amplitude instead of 
gradually decreasing tail obtained in the full calculation.    
The reason is that a single $k$ configuration for the $^3D$ channel 
is not enough and it gives an oscillating function 
of $\varphi^{(2)}_k(r)$ with the 
$e^{-\frac{\nu}{2}r^2} j_2(kr)$ dependence.  

The present analysis indicates that the superposition of different $k$
configurations in a 
wide momentum space is essential for detailed description of the 
tensor correlation, even though the contribution of $k\sim 2$ fm$^{-1}$ is dominant. 
In particular, higher $k$ components  in the $^3S$ and 
$^3D$ channels, typically $k\gtrsim 3$ fm$^{-1}$,  are necessary in precisely describing
the tensor correlation at shorter range around $r\sim 1$ fm of the $T=0$ pair.

\section{Results of ${}^8\textrm{Be}$}
In this section, we investigate the tensor correlations 
in $^8$Be with a two $\alpha$ configuration,
where the
V2m-3R interaction is adopted.
Here, AQCM-T
is applied to one of the $\alpha$ clusters,
and we 
adopt only two channels, $\beta=\{^3S,^3D\}$ with $k=\{1,2,3\}$ fm$^{-1}$. 

\begin{figure}[t]
\begin{center}
\includegraphics[width=8.5cm]{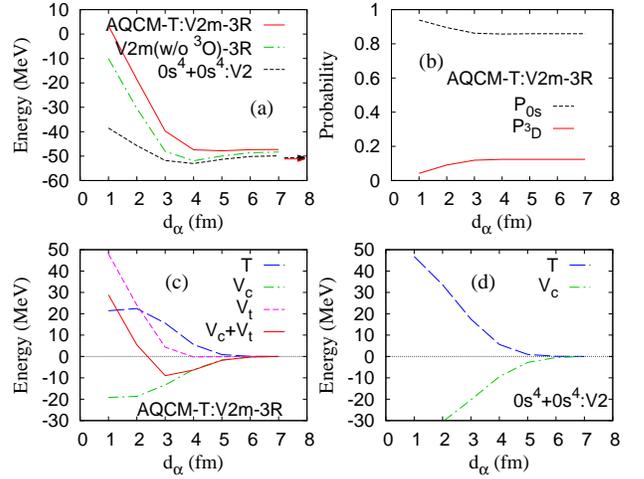} 
\end{center}
  \caption{	\label{fig:be8}
Energy 
and the probabilities of $0s$ and $D$-states in
$^8$Be 
calculated with AQCM-T and the V2m-3R interaction,
as a function of the relative distances $d_\alpha$
between the two $\alpha$ clusters,
 (a) total energy, (b) $0s$ and $^3D$ probabilities, 
and (c) energy of each component of the Hamiltonian. 
The total energy  
with the modified Majorana exchange term ($M=0.50$) is also shown in (a).
The results calculated 
with the BB cluster model and V2 are shown in 
(a) and (d). 
In panel (a), 
the asymptotic energy for AQCM-T and BB
are shown by arrows. 
The energy calculated with AQCM-T using the V2m central interaction without the odd part
combined with the 3-range fit tensor interaction
is also shown in (a).
In (c) and (d), the relative energies  measured from  $d_\alpha=7$ fm are plotted. 
}
\end{figure}

In Fig.~\ref{fig:be8}, we show the energies of $^8$Be as a function of
$d_\alpha$, which is the parameter for the relative distance between two $\alpha$ clusters,
(a): total energy, (b): probabilities of the $0s$ and $^3D$ configurations, 
(c): contribution of each component of the Hamiltonian.   
In (c) and (d), the relative energies are measured from values at $d_\alpha=7$ fm. 
Also we
list in Table~\ref{tab:be8}
the values for the contribution of each term of the Hamiltonian
and the probabilities of the $0s$ and $^3D$ configuration  
as functions of $d_\alpha$.
The results calculated with 
BB cluster model with V2 interaction are shown in 
Figs.~\ref{fig:be8} (a), (d), and Table~\ref{tab:be8}
for comparison.
In Fig.~\ref{fig:be8} (a) and Table~\ref{tab:be8}, we also show the ideal values of the asymptotic 
energies corresponding to the ones at
$d_\alpha\to \infty$, evaluated as twice the 
$^4\textrm{He}$ energy calculated with the consistent model space,
$i.e.$, the two-channel calculation with $k=\{1,2,3\}$ fm$^{-1}$. 
Note that 
here the constant shift of
$T_r=\hbar\omega/4$ 
(= $\hbar^2\nu/2m$, $m$ is the mean value of proton and neutron masses) 
is added for the kinetic and total energies,
which corresponds to the increase of the kinetic energy due to the
localization of the inter-cluster motion around $d_\alpha$.

In Table \ref{tab:be8},
we can confirm 
that
the two $\alpha$ system largely gains the 
tensor contribution
in the $d_\alpha\ge 6$ fm region
with significant mixing of 
$^3D$ and
dominant $0s$ components.
The energy of each component of the Hamiltonian in this region  is almost
comparable to the asymptotic values indicating that 
the two $\alpha$ system is approaching to a weak coupling
$^4\textrm{He}_\textrm{gs}+{}^4\textrm{He}_\textrm{gs}$ state. 
However, small deviations from the asymptotic values still remain,  
because, in the present model with the $\alpha_k+\alpha_0$ cluster wave functions,
higher order correlations of $\alpha_k+\alpha_k$ configurations,
where both clusters contain the correlated $T=0$ $NN$-pair, 
are omitted. 

As seen in Table~\ref{tab:be8} and Fig.~\ref{fig:be8} (a),
in the region of $d_\alpha\le 3$ fm, 
the system gets more excited as the $\alpha$-$\alpha$ distance
($d_\alpha$) becomes smaller. 
In particular, in the $d_\alpha\le 2$ fm region, 
the total energy rapidly increases, because the tensor correlation is remarkably suppressed
as can be seen in the reductions of $V_\mathrm{\tens}$ and $^3D$ in Table \ref{tab:be8} and Fig.~\ref{fig:be8} (b).  
Namely, although
the $V_\mathrm{\tens}$ contribution stays almost constant in the  $d_\alpha\ge 3$ fm region,
it rapidly decreases in the shorter region, the $d_\alpha\ge 2$ fm, as 
the $\alpha$ clusters come close to each other. 
Also the $^3D$ probability is almost unchanged in the
$d_\alpha\ge 3$ fm region, but it rapidly decreases in the $d_\alpha\le 2$ fm region. 
This means that the suppression of the tensor correlation strongly contributes to 
the repulsion between two $\alpha$ clusters at short distances. 
On the other hand, the total energy starts increasing already around  $d_\alpha\sim 3$ fm,
and its origin is the increase of the  kinetic energy 
rather than the tensor suppression. 
In other words, in the $\alpha$-$\alpha$ energy curve, 
the tensor suppression contributes to 
the repulsion at relatively short range between two $\alpha$ clusters,
whereas the increase of the kinetic energy contributes in the longer range.
These two repulsive effects enhance the characteristic development of the two $\alpha$ cluster structure in $^8\textrm{Be}$.

Both of these repulsive effects between two $\alpha$ clusters can be understood 
as the realization of the Pauli blocking effect, 
but they come from different origins. 
Indeed, the longer-range one
can be seen even in the BB calculation, because it comes 
from the increase of the kinetic energy due to the Pauli blocking 
between $0s$-orbit nucleons in the $\alpha$ cluster and that in the other $\alpha$ cluster.
However, the shorter-range one comes from the tensor suppression, 
which is the blocking of   
$(0s)^{-2}(0p)^{2}$ excitations induced by the tensor interaction in the correlated $\alpha$ cluster
by the other $\alpha$ cluster.
As discussed in the analysis for  $^4\textrm{He}$, the 
spatial extension of the tensor correlated
$NN$ pair is relatively smaller than the typical range of the uncorrelated $NN$ pair in the $(0s)^4$ state. 
As a result, the tensor suppression occurs 
only when two $\alpha$ clusters are close 
enough to block the particle hole excitation of the correlated pair
with a compact distribution.
This scenario of the tensor suppression and the consequent appearance 
of two $\alpha$ cluster structure
is consistent with the ones already proposed and discussed 
quite long time ago~\cite{Bando,Yamamoto}. 
We also discussed similar effect for the tetrahedron configuration of four $\alpha$ 
clusters in $^{16}$O that the finite distance between
$\alpha$ clusters is favored due to the tensor suppression~\cite{iSMT}.

As seen in Fig.~\ref{fig:be8} (a),
the BB calculation with V2 (dotted line) gives a shallow energy 
pocket around $d_\alpha=3\sim 4$ fm. 
However, in the present calculation with the V2m-3R interaction, 
this energy pocket disappears because of the weaker central interaction. 
If we remove the odd part of the central interaction,
we obtain an energy pocket with almost the same depth as the V2 interaction (see 
the dash-dotted line of Fig.~\ref{fig:be8} (a)). 
Note that this change of the odd part of the central interaction keeping the 
even part unchanged 
gives almost no effect to the $^4\textrm{He}$ results.
Fine tuning of the central interactions, in particular 
the odd part,  is an remaining problem for the study of heavier systems
in near future.

\begin{table}[ht]
\caption{Energy for each component of the Hamiltonian
and the probabilities of $0s$ and $D$-states in
$^8$Be 
calculated with AQCM-T and the V2m-3R interaction,
as a function of the relative distances $d_\alpha$
between the two $\alpha$ clusters (upper column).
The asymptotic and threshold energies are given 
as twice the 
$^4\textrm{He}$ energy calculated with a consistent model space.
For the asymptotic values, the 
constant shift of $T_r=\hbar\omega/4=5.2$ MeV is added
for the kinetic and total energies, 
corresponding to the localization of clusters with fixed relative distance.
The asymptotic value of the $0s$ probability given by the square of  
${\cal P}_{0s}$ for $^4\textrm{He}$ is also shown.
The results
obtained by the BB $2\alpha$ cluster model with the V2 interaction 
are also shown (lower column). 
\label{tab:be8}
}
\begin{center}
\begin{tabular}{cccccccc}
\hline
\hline
 \multicolumn{7}{c}{AQCM-T:V2m-3R} \\
$d_\alpha$  &	$\langle E\rangle_{2\alpha}$	&	$\langle T\rangle_{2\alpha}$&	
$\langle V_\mathrm{\cent}\rangle_{2\alpha}$	 &	$\langle V_\mathrm{\tens}\rangle_{2\alpha}$ &	${\cal P}_{0s}$	 & ${\cal P}_{^3D}$ \\ \hline
1 	&$	3.4 	$&$	148.1 	$&$	-132.8 	$&$	-16.1 	$&$	0.94 	$&$	0.04 	$\\
2 	&$	-18.6 	$&$	149.1 	$&$	-132.3 	$&$	-40.0 	$&$	0.90 	$&$	0.09 	$\\
3 	&$	-39.7 	$&$	142.3 	$&$	-127.0 	$&$	-59.8 	$&$	0.86 	$&$	0.12 	$\\
4 	&$	-47.4 	$&$	132.3 	$&$	-119.7 	$&$	-64.4 	$&$	0.86 	$&$	0.12 	$\\
5 	&$	-47.8 	$&$	127.6 	$&$	-115.3 	$&$	-64.2 	$&$	0.86 	$&$	0.12 	$\\
6 	&$	-47.4 	$&$	126.7 	$&$	-113.9 	$&$	-64.1 	$&$	0.86 	$&$	0.12 	$\\
7 	&$	-47.3 	$&$	126.7 	$&$	-113.7 	$&$	-64.1 	$&$	0.86 	$&$	0.12 	$\\
	&	$2\langle E\rangle_{\alpha}+T_r$	&	2$\langle T\rangle_{\alpha}+T_r$&	
$2\langle V_\mathrm{\cent}\rangle_{\alpha}$	 &	2$\langle V_\mathrm{\tens}\rangle_{\alpha}$ &  	
$\{ {\cal P}_{0s}\}^2_\alpha$ 	\\
asymp. 	&$	-51.2 	$&$	134.9 	$&$	-114.9 	$&$	-74.5 	$& 0.82 &$		$\\
$2\alpha$ thres.	&$	-56.4 	$&\\
\hline \hline
\multicolumn{7}{c}{BB:V2}\\
$d_\alpha$ (fm) &	$\langle E\rangle_{2\alpha}$&	$\langle T\rangle_{2\alpha}$		& $\langle V_\mathrm{\cent}\rangle_{2\alpha}$ \\ \hline					
1 	&$	-38.5 	$&$	145.3 	$&$	-187.6 	$\\						
2 	&$	-45.7 	$&$	132.1 	$&$	-181.4 	$\\						
3 	&$	-51.8 	$&$	116.0 	$&$	-171.1 	$\\						
4 	&$	-53.0 	$&$	104.1 	$&$	-160.2 	$\\						
5 	&$	-51.3 	$&$	99.5 	$&$	-153.5 	$\\						
6 	&$	-50.2 	$&$	98.6 	$&$	-151.3 	$\\						
7 	&$	-49.9 	$&$	98.5 	$&$	-150.8 	$\\	
&	$2\langle E\rangle_{\alpha}+T_r$	&	2$\langle T\rangle_{\alpha}+T_r$&	
$2\langle V_\mathrm{\cent}\rangle_{\alpha}$	 &	\\			
asymp. & $	-50.6 	$&$	98.5 	$&$	-150.7 	$&	\\	
$2\alpha$ thres.	&$	-55.8 	$&\\		
\hline	
\hline		
\end{tabular}
\end{center}
\end{table}

\section{Summary}

In this paper, we directly treated the tensor interaction and
examined the effect in $^4\textrm{He}$ and $^8\textrm{Be}$.
We extend the framework of $i$SMT and newly proposed AQCM-T, 
tensor version of AQCM.
Although the AQCM-T is a phenomenological model,
we can treat the $^3S$-$^3D$ coupling in the deuteron-like $T=0$ $NN$-pair induced 
by the tensor interaction
in a very simplified way,
which allows us to proceed to heavier nuclei.  
The model is also regarded  
a specific version of the HM-AMD.
In the previous analyses based on $i$SMT and HM-AMD,
the tensor interaction was just added to the effective Hamiltonian,
and the tensor effect was doubly counted.
In this study, we proposed a new effective 
interaction, V2m, 
where the triplet-even part of the central interaction (V2) was
reduced to 60\% of the original strength so as to reproduce the correct
binding energy of $^4\textrm{He}$ 
within the AQCM-T model space. 
For the tensor term, G3RS interaction was adopted,
which was refitted using three Gaussians with a factor of $r^2$.
This combination of the central and tensor interactions
is called V2m-3R.

For $^4\textrm{He}$, the two results of 
AQCM-T with V2m-3R and $(0s)^4$ configuration with V2 
give almost the same
total energy; however, 
the contributions of each component of the Hamiltonian are much different. 
The contribution of the central interaction is reduced by $\sim$20 MeV
in V2m-3R,
because of the weaker triplet-even channel compared with that in V2,
whereas the
large gain of the tensor energy compensates this reduction
and even overcomes the increase of the kinetic energy.
This effect is
attributed to the $D$-state mixing with the $S$-state, which is still a dominant component;
the $^3D$ probability (${\cal P}_{{}^3D}$) is only 8\%.
The AQCM-T calculation with V2m-3R gives the rms matter radius of 
$1.46$ fm, which well agrees with the value reduced from the
experimental charge radius, $1.455$ fm.

The present AQCM-T was also applied to $^8\textrm{Be}$
within the two $\alpha$ model space, where one of the $\alpha$ 
cluster was transformed from 
the $(0s)^4$ configuration using AQCM-T. The
tensor effect was investigated as a function of 
relative distance between $\alpha$ clusters, $d_\alpha$, 
and found to give significant contribution to the short-range repulsion 
between two $\alpha$ clusters. In the large $d_\alpha$ region, 
the contribution of each term of the Hamiltonian is
almost comparable to the asymptotic values deduced from 
twice of the values for $^4$He, indicating that 
the two $\alpha$ system is approaching to a weak coupling
$^4\textrm{He}_\textrm{gs}+{}^4\textrm{He}_\textrm{gs}$ state.
It also indicates that, although the present AQCM-T model for $^8\textrm{Be}$
explicitly treat the tensor correlation only in one of the $\alpha$ clusters, 
this is good approximation at least for the two $\alpha$ system 
owing to the Bosonic nature of the $\alpha$ clusters.

The tensor interaction is really the key ingredient of the cluster structure.
It contributes to the strong binding of the subsystems,
$^4\textrm{He}$ called $\alpha$ cluster, and
it is also related to the weak interaction 
between the subsystems.
Furthermore, the tensor suppression in each $\alpha$ cluster
contributes to the strong repulsion 
at short relative distances.
This scenario of the tensor suppression and the consequent appearance 
of two $\alpha$ cluster structure was proposed
quite long time ago, and here we have discussed it in a more direct way.
It is worthwhile to investigate such important effect 
of the tensor interaction in heavier nuclei,
which may be in capable because of the simple AQCM-T 
treatment  proposed here.
Extension of AQCM-T for nuclear matter is also an important issue, which may be associated with the 
saturation property of nuclear matter.

\appendix
\section{Gaussian fit for tensor force} \label{app:a}
\subsection{Determination of parameters for tensor interaction}
We aim to propose a G3RS-like tensor interaction in a convenient form, which can be easily used in practical
calculations; we fit the  G3RS tensor force with the multi-range Gaussian functional form with a factor of $r^2$ as follows.
\par
The radial part of the G3RS tensor term for the $^3E$ or $^3O$ channel is
\begin{align}
V_\mathrm{\tens}^{\mathrm{(G3RS)}\{^3E,^3O\}}(r)
=\sum_{n=1}^3V_{\mathrm{\tens},n}^{\mathrm{(G3RS)}\{^3E,^3O\}}\exp\left(-\frac{r^2}{\eta_{\mathrm{\tens},n}^2}\right),
\end{align}
and that of the $n_\mathrm{max}$-range fit introduced in this work is 
\begin{align}
V^{\mathrm{(fit)}\{^3E,^3O\}}_{\mathrm{\tens}}(r)
=\sum_{n=1}^{n_\mathrm{max}}V^{\mathrm{(fit)}\{^3E,^3O\}}_{\mathrm{\tens},n}r^2\exp\left(-\beta_nr^2\right).
\label{t-fit}
\end{align}
Our aim here is to fit $f(r)=V_\mathrm{\tens}^{\mathrm{(G3RS)}\{^3E,^3O\}}(r)$
with $g(r)=V_\mathrm{\tens}^{\mathrm{(fit)}\{^3E,^3O\}}(r)$. 
For this aim, we define the function $F$, which is the square of the difference
between these two integrated in the $a\le r\le b$ region as
\begin{align}
F(\{V^\mathrm{(fit)}_{\mathrm{\tens},n}\})=\int_a^b(f(r)-g(r))^2 dr.
\label{app.eq.F}
\end{align}
We optimize the strength parameters
$\{V^\mathrm{(fit)}_{\mathrm{\tens},n}\}$ 
by minimizing $F$, while the range
parameters $\{\beta_n\}$ and the  integration interval ($a$ and $b$) are fixed.
We construct the 3-range fit ($n_\mathrm{max}=3$) tensor interaction 
for the use of economical calculations.
For the range parameters  $\{\beta_n\}$ $(n=1,2,3)$  
of the $3$-range fit, we employ the values of Furutani 
interaction, whose functional form is identical to $g(r)$.
The range parameters $\{\beta_n\}$ $(n=1,2,3)$ are well scattered, so it is appropriate 
to take $a=1/\sqrt{\beta_3}$ and $b=1/\sqrt{\beta_1}$ in Eq.~\eqref{app.eq.F}.

As shown in Fig.~\ref{app.fig.TensorCompare}, 
the $r$-dependence of the 3-range fit  
well agrees
with that of the G3RS tensor part, but the fitting precision is not perfect.
We also prepare the 20-range fit
($n_\mathrm{max}=20$) version, which fits almost perfectly.
In order to reproduce the G3RS tensor part in a wide range, we choose the range parameters
$b_n=b_1\times(b_{20}/b_1)^{(n-1)/19}$ with $b_1=0.1\,\mathrm{fm}$ and $b_{20}=20.0\,\mathrm{fm}$, 
where $\beta_n=1/b_n^2$.
The parameters $\{b_n\}$ and $\{V^\textrm{(fit)}_{\mathrm{\tens},n}\}$ 
of the $20$-range fit are shown in Table~\ref{app.tab.20RangeFitParameter}.

\begin{table}[ht]
\caption{The parameters $\{b_n\}$ and $\{V^\textrm{(fit)}_{\mathrm{\tens},n}\}$
of the $20$-range fit
tensor interaction for the $^3E$ and 
$^3O$ channels.
}
\label{app.tab.20RangeFitParameter}
\centering
\begin{tabular}{rrrr}
\hline \hline
$n$ & $b_n\,\mathrm{(fm)}$ & \multicolumn{2}{c}{$V^\textrm{(fit)}_{\mathrm{\tens},n}$ $\mathrm{(MeV\cdot{}fm^{-2})}$}  \\
 &  &  $^3E$ & $^3O$ \\
 \hline
  $1$ &   $0.1000$ & $-1403.1$ & $467.67$ \\
  $2$ &   $0.1322$ & $387.27$ & $-129.09$ \\
  $3$ &   $0.1747$ & $-443.05$ & $147.72$ \\
  $4$ &   $0.2308$ & $50.189$ & $-16.847$ \\
  $5$ &   $0.3051$ & $-90.076$ & $30.307$ \\
  $6$ &   $0.4032$ & $-35.449$ & $10.705$ \\
  $7$ &   $0.5329$ & $-160.83$ & $48.483$ \\
  $8$ &   $0.7043$ & $-76.453$ & $22.860$ \\
  $9$ &   $0.9308$ & $-53.823$ & $16.165$ \\
$10$ &   $1.2302$ & $-12.189$ & $3.6997$ \\
$11$ &   $1.6258$ & $-0.73858$ & $0.28279$ \\
$12$ &   $2.1487$ & $-1.0741$ & $0.34869$ \\
$13$ &   $2.8398$ & $1.2481\times10^{-2}$ & $-1.2718\times10^{-3}$ \\
$14$ &   $3.7531$ & $-1.3513\times10^{-2}$ & $3.5475\times10^{-3}$ \\
$15$ &   $4.9602$ & $5.0273\times10^{-3}$ & $-1.3535\times10^{-3}$ \\
$16$ &   $6.5555$ & $-1.7096\times10^{-3}$ & $4.6338\times10^{-4}$ \\
$17$ &   $8.6638$ & $5.3737\times10^{-4}$ & $-1.4603\times10^{-4}$ \\
$18$ & $11.4503$ & $-1.4765\times10^{-4}$ & $4.0172\times10^{-5}$ \\
$19$ & $15.1329$ & $3.1296\times10^{-5}$ & $-8.5195\times10^{-6}$ \\
$20$ & $20.0000$ & $-3.7491\times10^{-6}$ & $1.0209\times10^{-6}$ \\ \hline\hline
\end{tabular}
\end{table}

\subsection{Precision of the fitting}
To evaluate the
precision of the fitting for the radial part of the tensor interactions $V^{\mathrm{(fit)}\{^3E,^3O\}}_{\mathrm{\tens}}(r)$,  
we calculate the matrix element
\begin{align}
\mathcal{F}=\langle\psi_f\mid\hat{V}_\mathrm{\tens}\mid\psi_i\rangle,
\label{app.eq.mathcalF}
\end{align}
where $\psi_{i,f}$ is normalized relative wave function for the $NN$ pair.
For the $^3E$ channel, 
the dominant contribution to the energy of $^4\textrm{He}$  
comes from the non-diagonal matrix element for the $S$-$D$ coupling 
between the $^3E$ pair in $(0s)^4$  
and the $^3D$ pair with $k=2$ $\mathrm{fm}^{-1}$
is especially important. 
The corresponding matrix element is calculated as 
\begin{align}
\mathcal{F}_{S\textrm{-}D}^{^3E}=&\sqrt{8}\int_0^\infty\varphi^{(2)}_k(r)V_\mathrm{\tens}^{^3E}(r)\varphi^{(0)}_{k=0}(r)r^2dr \nonumber \\
\propto&\sqrt{8}\int_0^\infty{}V_\mathrm{\tens}^{^3E}(r)e^{-\frac{\nu}{2}r^2}\left[e^{-\frac{\nu}{2}r^2}j_2(kr)\right]r^2dr,
\end{align}
where 
$\varphi^{(l)}_k(r)$ is the normalized radial wave function in Eq.~\eqref{eq:partial-NN} and proportional to 
$e^{-\frac{\nu}{2}r^2}j_l(kr)$.
For the $^3O$ channel, 
we evaluate here the diagonal matrix element for the $^3P$ pair 
contained in the $D$-state of $^4\textrm{He}$ 
using the ADCM-T wave function with $k=2$ $\mathrm{fm}^{-1}$ as 
\begin{align}
\mathcal{F}_{P\textrm{-}P}^{^3O}=&2\int_0^\infty\varphi^{(1)}_k(r)V_\mathrm{\tens}^{^3O}(r)\varphi^{(1)}_k(r)r^2dr \nonumber \\
\propto &2\int_0^\infty{}V_\mathrm{\tens}^{^3O}(r)\left[e^{-\frac{\nu}{2}r^2}j_1\left(\frac{k}{2}r\right)\right]^2r^2dr.
\end{align}
The calculated values of the $\mathcal{F}_{S\textrm{-}D}^{^3E}$ and 
$\mathcal{F}_{P\textrm{-}P}^{^3O}$ for the G3RS, Furutani, $3$-range fit, and $20$-range fit tensor interactions are shown in Table~\ref{app.tab.mathcalF}.
The values for the 3-range fit well agrees to those for G3RS within 
the accuracy of a few \%.
For the 20-range fit, the agreement is almost perfect meaning that it can be 
regarded as an equivalent potential to the G3RS tensor interaction.
\begin{table}[ht]
\caption{The values of the $\mathcal{F}_{S\textrm{-}D}^{^3E}$ and $\mathcal{F}_{P\textrm{-}P}^{^3O}$ 
calculated 
with the G3RS, Furutani, $3$-range fit, and $20$-range fit tensor interactions.}
\label{app.tab.mathcalF}
\centering
\begin{tabular}{rrr}
\hline \hline
$V^{\mathrm{(fit)}\{^3E,^3O\}}_{\mathrm{\tens}}(r)$ & $\mathcal{F}_{S\textrm{-}D}^{^3E}\,(\mathrm{MeV})$ & $\mathcal{F}_{P\textrm{-}P}^{^3O}\,(\mathrm{MeV})$ \\ \hline
G3RS & $-35.056$ & $6.8319$ \\
Furutani & $-44.980$ & $3.6135$ \\
3-range & $-34.805$ & $6.6619$ \\
20-range & $-35.058$ & $6.8323$ \\ \hline\hline
\end{tabular}
\end{table}


\begin{acknowledgments}
The computational calculations of this work were performed by using the
supercomputer in the Yukawa Institute for theoretical physics, Kyoto University. This work was supported by 
JSPS KAKENHI Grant Numbers  26400270 (Y.~K-E.), 17K05440 (N.~I.), 18J13400 (H.~M.), and 18K03617 (Y.~K-E.).
\end{acknowledgments}


\begin{references}

\bibitem{Brink}
D.~M.~Brink, in {\it Proceedings of the International School of Physics ``Enrico Fermi'' Course XXXVI},
edited by C.~Bloch (Academic, New York, 1966), p. 247.

\bibitem{Fujiwara}
Y.~Fujiwara {\it et al.,}
Prog. Theor. Phys. Suppl. \textbf{68}, 29 (1980).

\bibitem{Hoyle}
F.~Hoyle, D.~N.~F.~Dunbar, W.~A.~Wenzel, W.~Whaling, Phys. Rev. {\bf 92}, 1095c (1953).

\bibitem{Uegaki}
E.~Uegaki, S.~Okabe, Y.~Abe, and H.~Tanaka, Prog. Theor. Phys. {\bf 57}, 1262 (1977).

\bibitem{THSR}
A.~Tohsaki, H.~Horiuchi, P.~Schuck, and G.~R\"{o}pke, Phys.
Rev. Lett. {\bf 87}, 192501 (2001).

\bibitem{Maris}
P.~Maris, J.~P. Vary, and A.~M. Shirokov, Phys. Rev. C {\bf 79}, 014308 (2009).

\bibitem{Dreyfuss}
  A.~C.~Dreyfuss, K.~D.~Launey, T.~Dytrych, J.~P.~Draayer, and C.~Bahri,
  Phys.\ Lett.\ B {\bf 727}, 511 (2013).


\bibitem{Yoshida}
  T.~Yoshida, N.~Shimizu, T.~Abe, and T.~Otsuka,
  J.\ Phys.\ Conf.\ Ser.\  {\bf 569},  012063 (2014).

\bibitem{Shimodaya}
I.~Shimodaya, R.~Tamagaki, and H.~Tanaka, 
Prog. Theor. Phys. {\bf 27}, 793 (1962).



\bibitem{ATMS}
M.~Sakai, I.~Shimodaya, Y.~Akaishi, J.~Hiura, and H.~Tanaka,
Prog. Theor. Phys. Suppl. {\bf 56}, 32 (1974).

\bibitem{Kamada}
H.~Kamada {\it et al}., Phys. Rev. C {\bf 64}, 044001 (2001).


\bibitem{Bando}
H.~Band\=o, S.~Nagata, and Y.~Yamamoto, Prog. Theor. Phys. {\bf 44}, 646
(1970). 

\bibitem{QMC} 
R.~B.~Wiringa, Steven~C.~Pieper, J.~Carlson, and V.~R.~Pandharipande,
Phys. Rev. C {\bf 62}, 014001 (2000).

\bibitem{Yamamoto}
Y.~Yamamoto, T.~Togashi, and K.~Kat\=o,
Prog. Theor. Phys. {\bf 124}, 315 (2010).


\bibitem{KanadaEnyo:1995tb}
  Y.~Kanada-En'yo, H.~Horiuchi, and A.~Ono,
  Phys.\ Rev.\  C {\bf 52}, 628  (1995).

\bibitem{KanadaEnyo:1995ir}
  Y.~Kanada-En'yo and H.~Horiuchi,
  Phys.\ Rev.\  C {\bf 52}, 647 (1995).

\bibitem{AMDsupp} 
Y. Kanada-En'yo and H. Horiuchi,
Prog. Theor. Phys. Suppl. {\bf 142},  205 (2001).

\bibitem{KanadaEn'yo:2012bj}
  Y.~Kanada-En'yo, M.~Kimura, and A.~Ono,
 Prog. Theor. Exp. Phys. {\bf 2012},  01A202 (2012).

\bibitem{Neff}
T.~Neff and H.~Feldmeier, Nucl. Phys. \textbf{A738}, 357 (2004).

\bibitem{Roth}
R.~Roth, T.~Neff, and H.~Feldmeier, Prog. Part. Nucl. Phys. {\bf 65}, 50 (2010).

\bibitem{Chernykh}
M.~Chernykh, H.~Feldmeier, T.~Neff, P.~von~Neumann-Cosel, and A.~Richter,
Phys. Rev. Lett.  {\bf 98}, 032501 (2007); {\it ibid}. {\bf 105}, 022501 (2010).

\bibitem{Dote:2005un} 
  A.~Dot\'e, Y.~Kanada-En'yo, H.~Horiuchi, Y.~Akaishi, and K.~Ikeda,
  Prog.\ Theor.\ Phys.\  {\bf 115}, 1069 (2006).

\bibitem{Simple}
	N.~Itagaki, H.~Masui, M.~Ito, and S.~Aoyama, Phys. Rev. C {\bf 71}, 064307 (2005). 

\bibitem{Masui}
	H.~Masui and N.~Itagaki, Phys. Rev. C {\bf 75}, 054309 (2007). 

\bibitem{Yoshida2}
	T.~Yoshida, N.~Itagaki, and T.~Otsuka, Phys. Rev. C {\bf 79}, 034308 (2009).

\bibitem{Ne-Mg}
	N.~Itagaki, J.~Cseh, and M.~P{\l}oszajczak, Phys. Rev. C {\bf 83}, 014302 (2011).

\bibitem{Suhara}
  T.~Suhara, N.~Itagaki, J.~Cseh, and M.~P{\l}oszajczak,
Phys. Rev. C {\bf 87}, 054334 (2013).

\bibitem{Suhara2015}
  T.~Suhara and Y.~Kanada-En'yo,
  Phys.\ Rev.\ C {\bf 91},  024315 (2015).

\bibitem{Itagaki}
N.~Itagaki, H.~Matsuno, and T.~Suhara, Prog. Theor. Exp. Phys. {\bf 2016}, 
093D01 (2016). 

\bibitem{Itagaki-CO}
N.~Itagaki, Phys. Rev. C {\bf  94}, 064324  (2016). 

\bibitem{Matsuno}
H.~Matsuno, N.~Itagaki, T.~Ichikawa, Y.~Yoshida, and Y.~Kanada-En'yo,
Prog. Theor. Exp. Phys. {\bf 2017}, 063D01  (2017). 
 
\bibitem{Matsuno2}
H.~Matsuno and N.~Itagaki,
Prog. Theor. Exp. Phys. {\bf 2017}, 123D05 (2017).

\bibitem{O24}
N.~Itagaki and A.~Tohsaki,
Phys. Rev. C {\bf 97}, 014307 (2018). 

\bibitem{Otsuka}
T.~Otsuka, T.~Suzuki, R.~Fujimoto, H.~Grawe, and Y.~Akaishi,
Phys. Rev. Lett. {\bf 95}, 232502 (2005).

\bibitem{Itagaki-SMT}
N.~Itagaki, H.~Masui, M.~Ito, S.~Aoyama, and K.~Ikeda,
Phys. Rev. C {\bf 73}, 034310 (2006).

\bibitem{TOSM}
T.~Myo, K.~Kat\=o, and K.~Ikeda, Prog. Theor. Phys. {\bf 113}, 763 (2005).

\bibitem{TOSM2007}
T.~Myo, S.~Sugimoto, K.~Kat\=o, H.~Toki, and K.~Ikeda, Prog. Theor. Phys. {\bf 117}, 257 (2007).

\bibitem{TOSM2009}
T.~Myo, H.~Toki, and K.~Ikeda, Prog. Theor. Phys. {\bf 121}, 511 (2009).

\bibitem{TOSM2011}
T. Myo, A. Umeya, H. Toki, and K. Ikeda, Phys. Rev. C {\bf 84}, 034315 (2011).

\bibitem{TOSM2012}
  T.~Myo, A.~Umeya, H.~Toki\ForestGreen{,} and K.~Ikeda,
  Phys.\ Rev.\ C {\bf 86}, 024318 (2012).

\bibitem{Myo:2015rbv} 
  T.~Myo, H.~Toki, K.~Ikeda, H.~Horiuchi, and T.~Suhara,
Prog. Theor. Exp. Phys. {\bf 2015}, 073D02 (2015).
\bibitem{Myo:2017gjc} 
  T.~Myo, H.~Toki, K.~Ikeda, H.~Horiuchi, and T.~Suhara,
  Phys.\ Lett.\ B {\bf 769}, 213 (2017).




\bibitem{iSMT}
N.~Itagaki and A.~Tohsaki,
Phys. Rev. C {\bf 97}, 014304 (2018). 

\bibitem{Myo:2017amv}  
  T.~Myo, H.~Toki, K.~Ikeda, H.~Horiuchi, T.~Suhara, M.~Lyu, M.~Isaka, and T.~Yamada,
Prog. Theor. Exp. Phys.  {\bf 2017}, 111D01 (2017).

\bibitem{Myo:2017nmz} 
  T.~Myo,
Prog. Theor. Exp. Phys.  {\bf 2018}, 031D01 (2018).

\bibitem{Horii:2011nf} 
  K.~Horii, H.~Toki, T.~Myo, and K.~Ikeda,
  Prog.\ Theor.\ Phys.\  {\bf 127}, 1019 (2012).

\bibitem{Volkov}
A.~B.~Volkov, Nucl. Phys. \textbf{74}, 33 (1965).

\bibitem{G3RS} 
R.~Tamagaki, Prog. Theor. Phys. {\bf 39}, 91 (1968).

\bibitem{Furutani}
H.~Furutani, H.~Horiuchi, and R.~Tamagaki, Prog. Theor. Phys. \textbf{62}, 981 (1979).

\bibitem{SCpn}
L.~Koester and W.~Nistler, Z. Phys. A {\bf 272}, 189 (1975).

\bibitem{SCnn}
G.~F.~de~T\'eramond and B.~Gabioud,
Phys. Rev. C {\bf 36}, 691 (1987).

\bibitem{angeli13}
I.~Angeli and K.~P.~Marinova, At.~Data Nucl.~Data Tables {\bf 99}, 69 (2013).   



\end{references}
\end{document}